\documentclass[]{imsart}
\RequirePackage[OT1]{fontenc}
\RequirePackage{amsthm,amsmath}
\RequirePackage[numbers]{natbib}
\newcommand\asp{\hspace{4mm}}

\newcommand{\Rzero}{{R_0}}

\newcommand{\immigration}{\iota}

\newcommand{\meanBeta}{{\bar{\beta}}}
\newcommand\mystretch{\rule[-2mm]{0mm}{5mm} }

\renewcommand{\theenumii}{\theenumi.\arabic{enumii}}
\usepackage{algorithm}
\usepackage{graphicx}
\usepackage{mathrsfs}
\usepackage{amsfonts}
\usepackage{threeparttable,booktabs}
\usepackage{enumerate}
\usepackage{tikz}
\usepackage{listings}
\usepackage{multirow}
\usepackage{amsfonts}
\usepackage{amssymb}
\usepackage{bbm}
\usepackage{arydshln}
\newcommand\secintegrated{S1}
\newcommand\secproofs{S2}

\allowdisplaybreaks

\makeatletter
\def\thm@space@setup{%
  \thm@preskip=0.45cm
  \thm@postskip=\thm@preskip 
}
\makeatother

\startlocaldefs
\numberwithin{equation}{section}
\theoremstyle{plain}
\newtheorem{thm}{Theorem}[section]
\newtheorem{definition}[thm]{Definition}

\newtheorem{example}[thm]{Example}
\newtheorem{proposition}[thm]{Proposition}

\theoremstyle{remark}

\endlocaldefs

\newcommand{\Var}{\mathrm{Var}}
\newcommand{\X}{\mathbf{X}}

\newcommand{\Z}{\mathbf{Z}}
\newcommand{\N}{\mathbf{N}}
\newcommand{\x}{\mathbf{x}}
\newcommand{\y}{\mathbf{y}}
\newcommand{\z}{\mathbf{z}}
\newcommand{\n}{\mathbf{n}}
\newcommand{\Cov}{\mathrm{Cov}}

\usepackage{enumitem}
\setlist{  
	listparindent=\parindent,
	parsep=0pt,
}

\usepackage{smartdiagram}

\usepackage{subfigure,sidecap}
\usepackage{wrapfig}
\usepackage{color}

\usepackage{filecontents}

\begin{document}

\begin{frontmatter}
\title{Systemic Infinitesimal Over-dispersion on Graphical Dynamic Models}
\runtitle{SIOD on Graphical Dynamic Models}

\begin{aug}
\author{\fnms{Ning} \snm{Ning}\thanksref{e1}\ead[label=e1,mark]{patning@tamu.edu}}
\and
\author{\fnms{Edward L.} \snm{Ionides}\ead[label=e2]{ionides@umich.edu}}

\address{Department of Statistics,
	Texas A\&M University.
	\printead{e1}}
\address{Department of Statistics,
University of Michigan, Ann Arbor.
\printead{e2}}

\runauthor{Ning and Ionides}


\end{aug}
\begin{abstract}
Stochastic models for collections of interacting populations have crucial roles in scientific fields such as epidemiology and ecology, yet the standard approach to extending an ordinary differential equation model to a Markov chain does not have sufficient flexibility in the mean-variance relationship to match data. To handle that, over-dispersed Markov chains have previously been constructed using gamma white noise on the rates. We develop new approaches using Dirichlet noise to construct collections of independent or dependent noise processes. This permits the modeling of high-frequency variation in transition rates both within and between the populations under study. Our theory is developed in a general framework of time-inhomogeneous Markov processes equipped with a graphical structure, for which ecological and epidemiological models provide motivating examples. We demonstrate our approach on a widely analyzed measles dataset, adding Dirichlet noise to a classical SEIR (Susceptible-Exposed-Infected-Recovered) model. Our methodology shows improved statistical fit measured by log-likelihood and provides new insights into the dynamics of this biological system.
\end{abstract}
\begin{keyword}
\kwd{Directed graph} 
\kwd{Time-inhomogeneous stochastic processes} 
\kwd{Extra-demographic stochasticity} 
\kwd{Simultaneous jumps} 
\kwd{Epidemiology}
\end{keyword}	

\end{frontmatter}


\section{Introduction}
\label{sec:Introduction}

In this section, we first give the background and motivations in Section \ref{sec:Background},
summarize our contributions in Section \ref{sec:contributions},
and demonstrate how to use our general algorithm on a practical application in Section \ref{sec:application},
followed by performance comparisons with that of \cite{breto2009time} in real data analysis on a widely analyzed measles dataset in Section \ref{sec:comparation}. The organization of the paper is provided in Section \ref{sec:Organization}.

\subsection{Background and motivations}
\label{sec:Background}

In this paper, we consider a general stochastic graphical dynamic model (GDM).
Recalling that a dynamic model is a process whose state varies with time, a GDM is an interacting dynamic model equipped with a graphical structure, where there is a process associated with each vertex whose state varies with time and the states of other vertices. GDMs have wide applications in demography, queueing theory, performance engineering, epidemiology, biology, and other areas, whose examples include stochastic compartmental models used in population dynamics. 
However, mechanistically-inspired models of probabilistic evolution frequently do not contain sufficient variability to adequately match real-world data and further flexibility is still often required (\cite{ramsay2017dynamic}, page $15$). This provides motivation to bestow the desired flexibility on Markov counting processes (MCPs), which are usually used as the building blocks of GDMs.
For example, in epidemiology, the conceptual, theoretical, and computational convenience of MCPs has led to their widespread use for modeling disease transmission processes with stochastic compartment models, such as the Susceptible-Exposed-Infected-Recovered (SEIR) model and its generalizations. 

When there is at most one event taking place in a sufficiently short period, a MCP is called simple otherwise it is called compound.
\citet{breto2011compound}  showed that infinitesimal dispersion is an equivalent mathematical terminology: a MCP is said to have infinitesimal equi-dispersion (IED) if and only if it is simple and a MCP is said to have infinitesimal over-dispersion (IOD) if and only if it is compound.
Using the ratio-formed formula of infinitesimal dispersion, the variance function divided by the mean function of the MCP in a sufficiently short period, IED (resp. IOD, infinitesimal under-dispersion (IUD)) holds if the ratio $=$ (resp. $>$, $<$) one.
For example, the Poisson process has IED. There are two classes of motivations for modeling IOD.
First, the process in question indeed has such occurrences, such as the ruin model in \cite{albrecher2017reinsurance} that allows for multiple insurance claims to occur simultaneously (a phenomenon known as \textit{clusters} in actuarial science terminology). Second, in data analysis, we may have multiple event times that are short compared to the scale of primary interest. For example, New York state   has published daily estimates of the number of total COVID-19 tests conducted at (possibly) different time within the day. 

By appending gamma noise to constant transition rates, \cite{breto2011compound} proposed an IOD generation approach based on simple MCPs. Progresses have been made, which include, but are not limited to, the following: \cite{breto2012infinitesimal} provided a multivariate extension for some univariate MCPs considered in \cite{breto2011compound}  with time-homogeneous transition rate functions (TRFs); \cite{zhang2016characterizations} gave characterizations of discrete compound Poisson distributions and an application in probabilistic number theory; \cite{sendova2018poisson} introduced a compound Poisson counting process with logarithmic compounded distribution; \cite{li2020surplus} proposed a surplus process involving a compound Poisson counting process; 
the concept of simultaneous co-jumps was proposed in \cite{breto2021co} with time-homogeneous TRFs; \cite{gao2022applications} proposed a generalization of the classical compound Poisson model with claim sizes following a compound distribution. There is also similar interest in queueing theory, for instance, the batch Markovian arrival process, which extends the Markovian arrival process by allowing multiple events to occur simultaneously (see, e.g.,
\cite{maraghi2009batch, jayaraman2010batch}).


The assumption of constant or time-homogeneous transition rates is often unrealistic \citep{krak2017efficient}. Based on the
fundamental time-inhomogeneous birth process discussed in Chapter $7$ of \cite{klugman2013loss}, a time-inhomogeneous compound-birth process was recently proposed by \cite{sendova2020introducing}.
A further generalization from being time-inhomogeneous, is allowing the TRF of one MCP to also depend on the state of others, which is called interacting particle systems in mathematics terminology. In finance, the price of one asset usually depend on time and prices of other assets (e.g. equation $(2.7)$ of \cite{ding2021markov} and equation $(2.1)$ of \cite{ning2021well}).
In epidemiology, the TRF of one compartment in stochastic compartment models usually depend on time and states of other compartments, for example, the rate of new infections in the SEIR-typed Markov chain model (equation $(6)$ on page $332$ of \cite{breto2009time} and equation \eqref{eq:muSE} of this paper).
With this kind of general TRFs, \cite{breto2009time} developed the first over-dispersion methodology for real epidemic data fitting, and this approach has been widely used.
Although their TRFs are quite general, their theoretical foundation is limited to the IOD generation approach on constant rates proposed in \cite{breto2011compound}.

The long-standing gap between the models used in practice and the theory provided by \cite{breto2009time} and \cite{breto2011compound} is hard to fill directly, since it is theoretically challenging to know the detailed properties of a stochastic integral generated by a gamma process with a general function integrand.
Two natural questions arise: Can an algorithmic approach be developed that performs comparably or better than that of \cite{breto2009time} without a theory-practice gap? Is this new approach applicable in practice and compatible with modern likelihood-based inference methodologies (e.g. \cite{ionides2006inference,ionides2015inference,king2008inapparent}) to fully replace that of \cite{breto2009time}?
Graphs, as a kind of data structure that models a set of objects
(nodes) and their relationships (edges), can be used as a denotation of a large number of systems across
various areas. Because of their great expressive power, researches on analyzing GDMs systemically have been receiving more and
more attentions in many areas, 
\cite{chen2017network} on network reconstruction from high-dimensional ordinary differential equations,  \cite{katzfuss2020ensemble} on ensemble {K}alman methods for high-dimensional hierarchical dynamic space-time models,
\cite{ionides2020bagged} on bagged filters for partially observed spatiotemporal systems, \cite{ning2022iterated} on high-dimensional spatiotemporal online learning on large graphs, to name a few. 
Then two more fundamental questions arise directly: Can a systemic theory be defined properly and established rigorously on a general graph instead of merely on edges? What sort of mathematical tools are needed to build a systemic theory to be exploited algorithmically?

In this paper, we aim to address the above four questions. We will presently provide
a summary of our contributions in the next subsection. Then much of the remainder of
this section will be devoted to concretizing these summarizations through a practical application and algorithmic performance comparisons.

\subsection{Our contributions}

\label{sec:contributions}
In this paper, we define Systemic IOD (SIOD) for general GDMs, provide corresponding methodologies for general dynamics, generate associated general algorithms, and demonstrate the algorithmic performance on a benchmark epidemiological modeling challenge.
In sum, the contributions of this paper are four-fold:
\begin{enumerate}
	\item \textbf{General GDM and systemic definitions.} The GDM under consideration is general in terms of a general graph structure and general dynamics over it. 
	We consider TRFs as general positive functions of time and the state of the whole graph, while all the preceding IOD theoretical literature considered either constants or functions of time only. We focus on dynamics over a general directed graph, while all the preceding IOD theoretical literature worked on dynamics over a single arrow of the graph. We hence give appropriate definitions of systemic infinitesimal dispersion (SID), which are consistent with preceding literature locally with respect to a single arrow. Globally, our definitions allow users to flexibly add IOD to dynamics over some subgraphs while keeping dynamics over the rest subgraphs having IED. 
	\smallskip
	
	
	\item \textbf{Innovative methodogies and algorithms.} On one hand, there are occasions in practice that are appropriate to model with bounded processes, for example when modeling biological population counts. Hence,  under boundedness constraints, in Section \ref{sec:iod_bounded} we generate IOD using multinomial distributions,  over outgoing arrows with the same tail. An algorithmic Euler realization of the resulting Proposition \ref{prop:iod_multinomial}  is provided, which is a general algorithm (Algorithm \ref{fig:euler_beta}) for generating dynamics having IOD over connected outgoing arrows. Its application to a well-known case study in epidemiology is provided in Algorithm \ref{fig:euler_beta_measles}. On the other hand, unbounded processes have wide applications, such as the pure birth process.
	Thus, without boundedness constraints, in Section \ref{sec:iod_unbounded} 
	we propose a methodology for generating IOD  using negative multinomial distributions, over incoming arrows with the same head and a corresponding general algorithm can be developed analogously. 
	\smallskip
	
	\item \textbf{Wide applicability.}
	Our theoretical framework is sufficiently general to cover many situations, yet it also has various features that make it applicable in practice.
	First, only a weak assumption is required (existence of the second moment of a single dynamic), which is usually satisfied in practice; second, software implementation  using Dirichlet random variables is routine and computationally convenient; third, with our definition of SIOD, users can flexibly choose those subgraphs that are appropriate to apply our algorithm for overdispersion; fourth, the convenience of simulation from the proposed algorithms enables likelihood-based data fitting using simulation-based algorithms, among which we demonstrated using iterated filtering \citep{ionides2011iterated,ionides2006inference,ionides2015inference,king2008inapparent}; fifth, our model can describe overdispersion using just one additional parameter and this parameter can be inferred using these aforementioned simulation-based algorithms (Section \ref{sec:comparation}).
	
	\smallskip

	\item \textbf{Improved data fitting with new insights.}  
	Although the algorithm proposed in \cite{breto2009time} has been widely used, besides the long-standing theory-practice gap, there are long-existing concerns about interpreting its results. \cite{he2010plug} applied that algorithm on a benchmark epidemiological modeling challenge. They obtained $R_0=56.8$, which is the basic reproduction number that is central in epidemiological theory. 
	To explain the surprisingly large $R_0$ value, \cite{he2010plug} gave detailed possible explanations on pages $276-278$ therein. In Section \ref{sec:comparation}, we conduct fair comparisons by applying our algorithm with the same data, same model setting, and same inference algorithm. We achieved better data fitting in terms of maximum log-likelihood (ML), and our ML estimation (MLE) of $R_0$ is $34.09$. Thus, our method not only has better fit to data but also provides a resolution of a previous discrepancy.
\end{enumerate}

\subsection{Application}
\label{sec:application}

\begin{figure}[htbp!]
	\centering
	\includegraphics[width=0.6\textwidth]{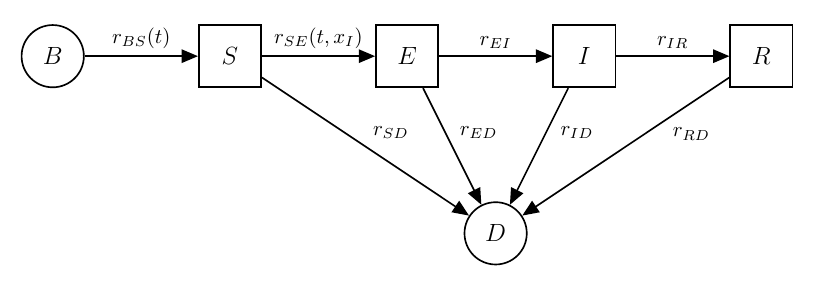}
	\caption{\footnotesize Directed graph for SEIR-type Markov chain models.}
	\label{measles_data_flowchart}
\end{figure}

In this subsection, we demonstrate our theories, methodologies, and algorithms through the measles application in \cite{he2010plug}.
Worldwide, measles remains a leading cause of vaccine-preventable death and disability, however global eradication of this highly infectious disease by intensive vaccination would be difficult. A fundamental class of models for measles transmission is the SEIR model, where
(S) represents susceptible individuals who have not been infected yet but may experience infection later, (E) represents individuals exposed and
carrying a latent infection,
(I) represents infectious individuals that have been infected and are infectious to others, and (R) represents recovered individuals that are no longer infectious and are immune. Two other compartments/vertices (B) and (D) representing the birth and death of individuals respectively, are added in SEIR-type Markov chain models which have been commonly used for measle data analysis. The directed graph in Fig. \ref{measles_data_flowchart} gives a diagrammatic representation, where arrows are used to indicate the possibility of transitions between vertices with labels parameterizing the transition rates.

The state of the system at time $t$ is given by the number of individuals in each vertex and is denoted as
$$\X(t)=\x=\{x_v\}_{v\in V}=\{x_B, x_S, x_E, x_I, x_R, x_D\},$$
where $V=\{B,S,E,I,R,D\}$. 
We use the standard definition of the transition rate
\begin{equation}
	\label{eqn:transition_rate_Q}
	Q(t,\x,\x'):=\lim_{h\downarrow 0}h^{-1}\mathbb{P}\big( \X(t+h)=\x' \mid \X(t)=\x \big),
\end{equation}
where $h$ is a sufficiently small time increment.
The standard interpretation of Fig. \ref{measles_data_flowchart} as a Markov chain having transition rates, conditional on $\x$, is given by
\begin{align*}
	&Q(t,(x_B,x_S,\x_{V\backslash \{B,S\}}),(x_B-1,x_S+1, \x_{V\backslash \{B,S\}}))=r_{BS}(t)x_B\mathbbm{1}_{\{x_B>0\}},\\
	&Q(t,(x_S,x_E,\x_{V\backslash \{S,E\}}),(x_S-1,x_E+1, \x_{V\backslash \{S,E\}}))=r_{SE}(t,x_I)x_S\mathbbm{1}_{\{x_S>0\}},\\
	&Q(t,(x_E,x_I,\x_{V\backslash \{E,I\}}),(x_E-1,x_I+1, \x_{V\backslash \{E,I\}}))=r_{EI}x_E\mathbbm{1}_{\{x_E>0\}},\\
	&Q(t,(x_I,x_R,\x_{V\backslash \{I,R\}}),(x_I-1,x_R+1, \x_{V\backslash \{I,R\}}))=r_{IR}x_I\mathbbm{1}_{\{x_I>0\}},\\
	&Q(t,(x_S,x_D,\x_{V\backslash \{S,D\}}),(x_S-1,x_D+1, \x_{V\backslash \{S,D\}}))=r_{SD}x_S\mathbbm{1}_{\{x_S>0\}},\\
	&Q(t,(x_E,x_D,\x_{V\backslash \{E,D\}}),(x_E-1,x_D+1, \x_{V\backslash \{E,D\}}))=r_{ED}x_E\mathbbm{1}_{\{x_E>0\}},\\
	&Q(t,(x_R,x_D,\x_{V\backslash \{R,D\}}),(x_R-1,x_D+1, \x_{V\backslash \{R,D\}}))=r_{RD}x_R\mathbbm{1}_{\{x_R>0\}},
\end{align*}
The time-inhomogeneous transition rate over $(B,S)$, denoted as $r_{BS}(t)$, is the per-capita rate of recruitment of susceptibles depending on known birth rates obtained via interpolation from birth records. 
A cohort-entry effect is also considered in calculating $r_{BS}(t)$, to reflect the fact that a large cohort of first-year students enters the schools each fall: a fraction $\theta_{c}$ of recruits into the susceptible class enter on the school admission day and the remaining fraction ($1-\theta_{c}$) enter the susceptible class continuously.
We specify the force of infection as
\begin{equation}
	r_{SE}(t,x_I)=\beta(t)(x_I+\immigration)^{\alpha}/{N(t)},
	\label{eq:muSE}
\end{equation} 
where $\beta(t)$ is the transmission rate, $\iota$ describes imported infectives, 
$\alpha$ is a mixing parameter with $\alpha=1$ corresponding to homogeneous mixing, and 
$N(t)$ is a known population size obtained via interpolation from census data.
Since transmission rates are closely linked to contact rates among children, which are higher during school terms, $\beta(t)$ reflects the pattern of school terms and holidays, as follows:
\begin{equation}
	\beta(t)=\begin{cases}
		(1+2 \{1-p\}\theta_{a} )\, \meanBeta &\quad\quad \mbox{ during school term},\\
		( 1-2p\theta_{a}) \, \meanBeta&\quad\quad \mbox{ during vacation},
	\end{cases} \label{eq:term}
\end{equation}
where $p$ is the proportion of the year taken up by school term,
$\meanBeta$ is the mean transmission rate, and $\theta_{a}$ measures the relative effect of school holidays on transmission.
For ease of interpretation, $\meanBeta$ is reparameterized in terms of $\Rzero$ which is the annual average basic reproductive ratio, such that $R_0=\meanBeta/r_{IR}$, where $r_{IR}$ is the recovery rate.
Here, $r_{EI}$ is the rate at which exposed individuals become infectious and
$r_{SD}=r_{ED}=r_{ID}=r_{RD}$ denotes a constant per capita death rate.

\begin{algorithm}[!htb]
	\noindent\begin{tabular}{l}
		Set the initial value $\X(0)$ and time interval $[0,T]$.\\
		Set time increment $\delta=T/N$ for integer $N$; define $t_n=n\delta$.\\
		FOR $n=0 \mbox{ to } N-1$\\
		\asp Generate $\{\Pi_{S0}, \Pi_{SE}, \Pi_{SD}\}$  according to the Dirichlet distribution  \mystretch\\
		\asp \asp\asp  $\operatorname{Dir}(\{\alpha_{S0}, \alpha_{SE}, \alpha_{SD}\})$ having \mystretch\\
		\asp \asp\asp  $\alpha_{SE}=c\pi_{SE}, \quad\alpha_{SD}=c\pi_{SD},\;\text{and}\; \; \alpha_{S0}=c-\alpha_{SE}-\alpha_{SD},$\mystretch\\
		\asp \asp\asp   where \mystretch\\
		\asp \asp\asp  $\pi_{SE}=\left(1-e^{-\int_{t_n}^{t_{n+1}}r_{SE}(s,\x)ds-r_{SD}\delta}\right)\frac{r_{SE}(t_n,\x)}{r_{SE}(t_n,\x)+r_{SD}}$,	\mystretch\\	
		\asp \asp\asp  $\pi_{SD}=\left(1-e^{-\int_{t_n}^{t_{n+1}}r_{SE}(s,\x)ds-r_{SD}\delta}\right)\frac{r_{SD}}{r_{SE}(t_n,\x)+r_{SD}}$.	\mystretch\\
		\asp   Generate process increments \mystretch\\
		\asp \asp\asp   $\{\Delta^{\X}_{S0}, \Delta^{\X}_{SE}, \Delta^{\X}_{SD}\}\sim \mbox{Multinomial}(X_S(t_n),\{\Pi_{S0}, \Pi_{SE}, \Pi_{SD}\})$\mystretch\\
		\asp \asp\asp   where $\Delta^{\X}_{S0}$ stands for retain individuals.\mystretch\\
		\asp Set $X_S(t_{n+1})=\Delta^{\X}_{S0}+\Delta^{\X}_{BS}$\mystretch\\
		END FOR
	\end{tabular}
	\caption{Euler scheme on generating dynamics having IOD over arrows $(S,E)$ and $(S,D)$, using Algorithm \ref{fig:euler_beta}.}
	\label{fig:euler_beta_measles}
\end{algorithm}

Our transition rates are taken the same as \cite{he2010plug} for equidispersed arrows, therefore we use the same Euler approximation. 
The dynamics over $(B,S)$ is modelled as an inhomogeneous Poisson process on each step of the Euler scheme.
The dynamics over outgoing connected arrows $\{(S,E), (S,D)\}$ (resp. $\{(E,I), (E,D)\}$, $\{(I,R), (I,D)\}$) are modeled through multinomial distributions on each step of the Euler scheme, and the dynamic over $(R,D)$ can be implied through fixed population. The only difference between our approach and that of \cite{breto2009time}, is the modeling of dynamics over $\{(S,E), (S,D)\}$.
Algorithm \ref{fig:euler_beta_measles} is obtained by applying our general Algorithm \ref{fig:euler_beta} to this application.
In Algorithm \ref{fig:euler_beta_measles} the event probabilities in the multinomial distribution are Dirichlet random variables, whereas the approach of \cite{breto2009time} is adding gamma noise to those probabilities. By  Example \ref{example_0}, the dynamic over $(B,S)$ has IED. The equation  \eqref{eqn:multinomial_pdf_simplified2} shows that the dynamics over connected outgoing arrows can be modeled as the multinomial distribution, if the dynamic over each of these arrows is modeled by the cumulative death process. Hence, by Example \ref{example_0}, the dynamic over each of the arrows $ \{(E,I),(I,R),(E,D),(I,D),(R,D)\}$ has IED. Given that the dynamics over arrows $\{(S,E), (S,D)\}$ are generated by Algorithm \ref{fig:euler_beta_measles}, by Proposition \ref{prop:iod_multinomial}, they have IOD. Then by our definition of SID, Definition \ref{def:system_infinitesimal_dispersion}, which  says that SIOD holds if there exists at least one arrow over which its dynamic has IOD while the dynamics over all other arrows have IED, this SEIR-type Markov chain has SIOD.



\subsection{Comparison}
\label{sec:comparation}
\begin{figure}[htbp!]
	\centering
	\includegraphics[width=0.6\textwidth]{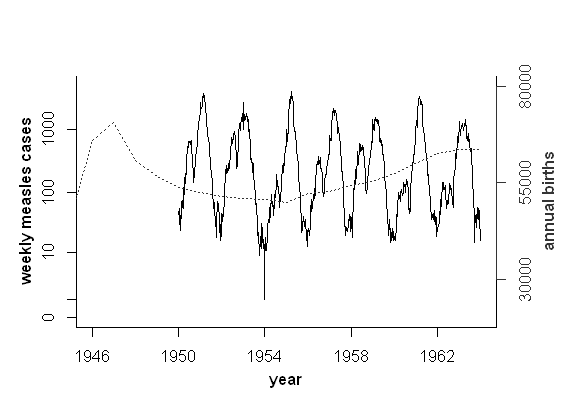}
	\caption{\footnotesize Weekly reported measles cases (solid line) and annual births (dotted line) for London}
	\label{fig:london_data}
\end{figure}

\cite{he2010plug} used the over-dispersion methodology (Box $1$ on page $280$ therein) proposed in \cite{breto2009time} on analyzing measles epidemics occurring in London during the pre-vaccination era, which is a well-tested and publicly accessible dataset with reported cases from $1950$ to $1964$. Figure \ref{fig:london_data} shows the case reports and annual birth rates for London. 
In order to conduct fair comparisons, we use the same model setting and data as \cite{he2010plug}. 
Thus, we fix $p=0.7589$ in \eqref{eq:term}, set the delay from birth to susceptible as $4$, and set the mortality rate $r_{SD}=1/50=0.02$ per year. 
The unknown model parameters in the SEIR-type Markov chain model covered in Section \ref{sec:application}, are $R_0$, $r_{EI}$, $r_{IR}$, $\alpha$, $\iota$,  $\theta_{c}$, and $\theta_{a}$. 
To calculate the likelihood of the data, a measurement model is added to describe the relationship between the latent disease dynamics and the observed case reports.
We use the same measurement model as \cite{he2010plug}, which has two more unknown parameters: reporting rate $\rho$ and dispersion parameter $\psi$ (see page $281$ of \cite{he2010plug} for a detailed description of this report measurement process).
The unknown initializations are $X_S(0)$, $X_E(0)$, $X_I(0)$, and $X_R(0)$. The unknown IOD model parameters are $\sigma_{SE}$ of the gamma noise-based approach used in \cite{he2010plug} and $c$ of our approach in Algorithm \ref{fig:euler_beta}. We implemented the same parameter inference algorithm (\cite{ionides2015inference}) as \cite{he2010plug}, via the pomp package (\cite{king2015statistical}). From Table \ref{table:Comparison_profile_MLE}, we can see that with the same number of unknown parameters which indicates the same complexity of inference, our method has better data fitting in terms of a higher ML.
\begin{table}[h!]
	\centering
	\begin{tabular}{||c c c| c c c||}
		\hline
		Name & \cite{he2010plug}  & Our method & Name & \cite{he2010plug}  & Our method \\ [0.5ex]
		\hline\hline
		ML & \textbf{-3804.9} & \textbf{-3803.2} & $\theta_{c}$ & 0.56 & 1 \\ 
		\textbf{$R_0$} & \textbf{56.8} & \textbf{34.09} & $\theta_{a}$ & 0.55 & 0.48 \\
		$r_{EI}$ & 28.9 & 52.71 &$X_S(0)$ & 0.0297 & 0.032 \\ 
		$r_{IR}$ & 30.4 & 22.88 &$X_E(0)$ & 5.17e-05 & 6.99e-05 \\ 
		$\alpha$ & 0.976 & 1.017 &$X_I(0)$ & 5.14e-05 & 4.52e-05 \\
		$\iota$ & 2.9 & 55.08 &$X_R(0)$ & 0.97 & 0.968 \\ 
		$\rho$ & 0.488 & 0.492 &$c$ & N/A & 652.8 \\ 
		$\psi$ & 0.116 & 0.118 &$\sigma_{SE}$ & 0.088 & N/A \\[1ex] 
		\hline
	\end{tabular}
	\caption{Comparisons of ML and MLEs}
	\label{table:Comparison_profile_MLE}
\end{table}

\begin{figure}[htbp!]
	\centering
	\includegraphics[width=9cm, height=6cm]{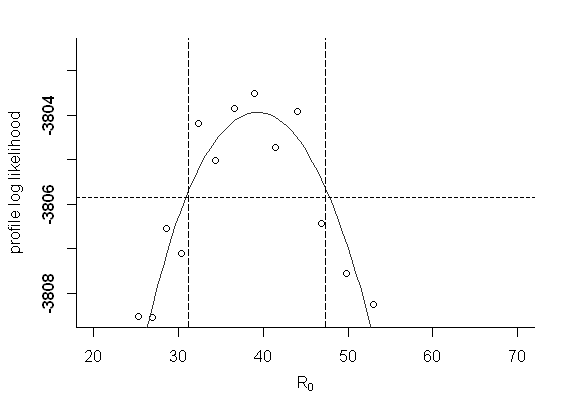}
	\caption{\footnotesize Log-likelihood analysis of the basic reproductive ratio, $R_0$. The dashed lines construct
		a $95\%$ confidence interval of $(31.21, 47.37)$ for London.}
	\label{fig:R0_profile}
\end{figure}

There are long-existing concerns about interpreting results generated by the approach proposed in \cite{breto2009time}.
The quantity $R_0$ is central in the epidemiological theory
because it has interpretations in terms of many quantities
of interest, which include mean age of the first infection,
mean susceptible fraction, exponential-phase epidemic
growth rate, and vaccination coverage required for eradication.
\cite{he2010plug} obtained MLE $R_0=56.8$ and the likelihoods over $R_0$ yielded a $95\%$ confidence interval of $(37, 60)$.
Furthermore, \cite{bjornstad2002dynamics} found an estimate of $R_0=29.9$ for London.
Hence, \cite{he2010plug} gave detailed possible explanations on pages $276-278$ therein, regarding concerns about the surprisingly high MLE value of $R_0$.
We obtained MLE $R_0=34.09$ and a $95\%$ confidence interval of $(31.21, 47.37)$ for London (Figure \ref{fig:R0_profile}).
Thus, our method not only shows improved statistical fit but also
provides a resolution of a previous discrepancy between continuous-time models fitted to time series data and other lines of evidence concerning $R_0$ for measles. 

\subsection{Organization of the paper}
\label{sec:Organization}
The rest of the paper proceeds as follows.
In Section \ref{sec:Model_Setup}, we give the graph structure, configurations on the graph, dynamics over the graph, and definitions of infinitesimal dispersion.  We illustrate our definitions with examples in Section {\secintegrated} of the supplementary material, through comparison with the definitions of integrated dispersion.
Our methodology for generating SIOD is provided in Section \ref{sec:Overdispersion}, where subections \ref{sec:iod_bounded} and \ref{sec:iod_unbounded} cover dynamics with and without boundedness constraints, respectively. Proofs of the paper are provided in Section  {\secproofs} of the supplementary material.
Code and data reproducing our results are available online at https://github.com/patning/Over-dispersion.  The notations used throughout this paper are listed in Table~\ref{tab:TableOfNotation}.

\section{Time-inhomogeneous GDMs}
\label{sec:Model_Setup}
In this section, we first give the graph structure and configurations on the graph in Section \ref{sec:Model_Setup_graph}, 
dynamics over graph and their transition rates in Section \ref{sec:Model_Setup_dynamics}, and then definitions of infinitesimal dispersion in
Section \ref{sec:Model_Setup_dispersion}.

\subsection{General directed graph}
\label{sec:Model_Setup_graph}
A directed graph is a set of vertices connected by edges, where each edge has a direction associated with it. In this paper, we consider a finite directed graph as $G = (V, A)$, where $V$ is a set of vertices and $A$ is a set of arrows. For the directed graph in Figure \ref{measles_data_flowchart},  $V=\{B,S,E,I,R,D\}$ and
$$A=\{(B,S),(S,E),(E,I),(I,R),(S,D),(E,D),(I,D),(R,D)\}.$$ 
For two vertices $v, v' \in V$, an arrow $(v, v')$ is considered to be directed from $v$ to $v'$; $v'$ is called the head and $v$ is called the tail of the arrow. In this paper, we allow the directed graph to have loops, i.e., arrows that directly connect vertices with themselves, while as in the typical setting we forbid self-loops, i.e. arrows of the form $(v, v)$ are not
contained in $G$. For a vertex $v \in V$, the number of head ends adjacent to $v$ is called the indegree of $v$ and is denoted as $\deg ^{-}(v)$; the number of tail ends adjacent to $v$ is called the outdegree of $v$ and is denoted as $\deg ^{+}(v)$. 
A vertex with zero indegree is called a source and the set of all source vertices is denoted by $S_o$. A vertex with zero outdegree is called a sink and the set of all sink vertices is denoted by $S_i$.
Thus,
$$S_o:=\{v\in V;\; \deg ^{-}(v) = 0\}\quad\text{and} \quad S_i:=\{v\in V;\; \deg ^{+}(v) = 0\}.$$
For the directed graph in Figure \ref{measles_data_flowchart}, $S_o=\{B\}$ and $S_i=\{D\}$.
Denote the set of incoming neighbors of $v$ as $N_G^{-}(v)$, which is the set of vertices $\overline{v}\in A$ such that  $(\overline{v}, v)\in A$. Denote the set of outgoing neighbors of $v$ as $N_G^{+}(v)$, which is the set of vertices $v'\in A$ such that  $(v, v')\in A$. Directed graphs distinguish between $N_G^{-}(v)$ and $N_G^{+}(v)$.

Given a Polish space $\mathcal{H}$, we let $\mathcal{D}_{\mathcal{H}}[0,\infty)$ denote the space of $\mathcal{H}$-valued c{\`a}dl{\`a}g functions on $[0,\infty)$, endowed with the Skorokhod $J_1$ topology, such that $\mathcal{D}_{\mathcal{H}}[0,\infty)$ is a Polish space; see \cite{parthasarathy2005probability} for further theoretical details. 
Denote the spin on any vertex $v\in V$ at any time $t\in [0,\infty)$ as $X_v(t)$. We consider that $X_v(t)$ is defined
on a probability space $(\Omega,\mathcal{F},\mathbb{P})$ and takes values in $\mathcal{D}_{\mathcal{H}}[0,\infty)$, equipped with the
Borel $\sigma$-algebra generated by open sets under the Skorokhod $J_1$-topology.
The number of transitions from vertex $v$ to vertex $v'$ through arrow $(v, v')\in A$ is modeled by a nondecreasing integer-valued jump process $N^{\X}_{vv'}(t)$ for $t\in [0,\infty)$ defined on the same probability space $(\Omega,\mathcal{F},\mathbb{P})$, where we use the customary initialization $N^{\X}_{vv'}(0)=0$. 

\subsection{Dynamics on a graph}
\label{sec:Model_Setup_dynamics}
Suppose that the dynamics of $\X(t):=\{X_v(t)\}_{v\in V}$ are driven by $\N^{\X}(t):=\{N^{\X}_{vv'}(t)\}_{(v, v')\in A}$ as follows: For $v\in V$ and $t\in [0,\infty)$
\begin{equation*}
	X_v(t)=X_v(0)+\sum_{\overline{v}\in N_G^{-}(v)}N^{\X}_{\overline{v}v}(t)-\sum_{v'\in N_G^{+}(v)}N^{\X}_{vv'}(t).
\end{equation*}
That is, the spin of vertex $v$ at time $t$ is given by its initial value at time $0$, plus the increments from all its incoming neighbors, and then minus the decrements to its outgoing neighbors. 
With respect to the whole graph, we define the gobal transition rate, as follows:
\begin{align*}
	q(t,\x,\mathbf{l}):=\lim_{h\downarrow 0}h^{-1}\mathbb{P}\big( \N^{\X}(t+h)=\n+\mathbf{l},\X(t+h)=\x+\mathbf{u} \mid \N^{\X}(t)=\n, \X(t)=\x \big),
\end{align*}
where $\mathbf{l}=\{l_{vv'}\}_{(v, v')\in A}$ and $\mathbf{u}=\{u_v\}_{v\in V}$ satisfy
$$u_v=\sum_{\overline{v}\in N_G^{-}(v)}l_{\overline{v}v}-\sum_{v'\in N_G^{+}(v)}l_{vv'},$$
and set other transition rates to zero.

Writting $A=\{{v_i'}{v_i''}\}_{i\in |A|}$ where $|A|$ stands for the cardinality of $A$, then with respect to arrows $\{{v_i'}{v_i''}\}_{i\in |A|}$, we define the arrow-based transition rate to measure simultaneously increments among arrows, as follows:
$$q_{\{{v_i'}{v_i''}\}_{i\in |A|}}(t,\x,(k_i)_{i\in |A|}):=\sum_{\mathbf{l}:\;\{l_{v_i'v_i''}=k_i\}_{i\in |A|}}q(t,\x,\mathbf{l}).$$ 
That is, $q_{\{{v_i'}{v_i''}\}_{i\in |A|}}(t,\x,(k_i)_{i\in |A|})$ is the summation of  transition rates with respect to arrow $({v_i'},{v_i''})\in A$ for each $i\in |A|$, such that $k_i$ units transfer simultaneously through arrow $(v_i',v_i'')\in A$. For example, if we are merely interested in the transition rate of a single arrow, say $(v_1',v_1'')$, 
$$q_{v_1'v_1''}(t,\x,k_1)=\sum_{\mathbf{l}:\;\{l_{v_1'v_1''}=k_1\}}q(t,\x,\mathbf{l}),$$
which is the transition rate that $k_1$ units transfer simultaneously from vertex $v_1'$ to vertex $v_1''$ through arrow $(v_1',v_1'')\in A$.
If we are interested in a pair of arrows, say $(v_1',v_1'')$ and $(v_2',v_2'')$ i.e., the case that $i$ in the set $\{1,2\}$, 
\begin{align*}
	q_{v_1'v_1'',v_2'v_2''}(t,\x,(k_1,k_2))=\sum_{\mathbf{l}:\;\{l_{v_1'v_1''}=k_1,\;l_{v_2'v_2''}=k_2 \}}q(t,\x,\mathbf{l}),
\end{align*}
which is the transition rate that simultaneously $k_1$ units transfer  from vertex $v_1'$ to vertex $v_1''$ through arrow $(v_1',v_1'')\in A$ and $k_2$ units transfer from vertex $v_2'$ to vertex $v_2''$ through arrow $(v_2', v_2'')\in A$.



\subsection{Measures of dispersion}
\label{sec:Model_Setup_dispersion}
Measures of dispersion were defined previously in the variance to mean ratio form (e.g. \cite{gillespie1984molecular}) and the variance and mean difference form (e.g. \cite{brown1998variance}). For theoretical analysis of dispersion, these two kinds of definitions are mainly equivalent while the difference-formed definition avoids the ``$0/0$" situation. However, the ratio-formed definition is widely used, partially due to the fact that it facilitates the dispersion comparison among different metrics and/or units. When it comes to data analysis, the over-dispersion parameter in Poisson regression (see, e.g., \cite{berk2008overdispersion}) uses the ratio-formed definition. In this paper, we properly define the SID with respect to the whole graph in Definition \ref{def:system_infinitesimal_dispersion}, which is the first time the measure of dispersion is defined on a graph to our best knowledge. Definition \ref{def:system_infinitesimal_dispersion} is formulated in terms of the measure of dispersion with respect to each arrow of the graph, whose definition is given below and is consistent with that in \cite{breto2011compound} (equation $(3)$ on page $2574$):
\begin{definition}
	\label{def:infinitesimal_dispersion}
	For arrow $(v,v')\in A$, define the infinitesimal variance 
	$$[\sigma_{vv'}^{d\X}(t,\x)]^2:=\lim_{h\downarrow 0}h^{-1}\Var[\Delta^{\X}_{vv'}(t,h)\mid \X(t)=\x],$$
	and the infinitesimal mean 
	$$\mu_{vv'}^{d\X}(t,\x):=\lim_{h\downarrow 0}h^{-1}\mathbb{E}[\Delta^{\X}_{vv'}(t,h)\mid \X(t)=\x],$$
	where 
	$$\Delta^{\X}_{vv'}(t,h):=N^{\X}_{vv'}(t+h)-N^{\X}_{vv'}(t).$$
	Define the infinitesimal dispersion index as the following ratio if it exists:
	$$D_{vv'}^{d\X}(t,\x):=[\sigma_{vv'}^{d\X}(t,\x)]^2 \Big/ \mu_{vv'}^{d\X}(t,\x).$$
	We say that with respect to arrow $(v,v')$,  $\X(t)$ has IED at $\X(t)=\x$ if $D_{vv'}^{d\X}(t,\x)=1$, has IOD at $\X(t)=\x$ if $D_{vv'}^{d\X}(t,\x)>1$, and has IUD at $\X(t)=\x$ if $D_{vv'}^{d\X}(t,\x)<1$.
\end{definition}

Noting that Definition \ref{def:infinitesimal_dispersion} is with respect to a specific arrow, now we give definitions with respect to the whole graph.  A GDM having SIOD was provided in Section \ref{sec:application}.
\begin{definition}
	\label{def:system_infinitesimal_dispersion} We say that 
	\begin{itemize}
		\item	$\X(t)$ has SIED at $\X(t)=\x$, if $D_{vv'}^{d\X}(t,\x)=1$ for all $(v,v')\in A$; 
		\item $\X(t)$ has SIOD at $\X(t)=\x$, if $D_{vv'}^{d\X}(t,\x)\geq 1$ for all $(v,v')\in A$ and there exists $(v_0,v_0')\in A$ such that $D_{v_0v_0'}^{d\X}(t,\x)> 1$; 
		\item  $\X(t)$ has SIUD at $\X(t)=\x$, if $D_{vv'}^{d\X}(t,\x)\leq 1$ for all $(v,v')\in A$ and there exists $(v_0,v_0')\in A$ such that $D_{v_0v_0'}^{d\X}(t,\x)< 1$. 
	\end{itemize}
\end{definition}

Note that the above definitions depend on arrow-wise variances.
To explore the infinitesimal correlations between two arrows' dynamics, in the following we give the pairwise definition of infinitesimal covariance consistently with the arrow-wise definition of infinitesimal variance in Definition \ref{def:infinitesimal_dispersion}.
\begin{definition}
	\label{def:infinitesimal_covariance}
	For arrows $(u,u')\in A$ and $(v,v')\in A$, define the infinitesimal covariance 
	$$\sigma_{uu',vv'}^{d\X}(t,\x):=\lim_{h\downarrow 0}h^{-1}\Cov[\Delta^{\X}_{uu'}(t,h),\Delta^{\X}_{vv'}(t,h)\mid  \X(t)=\x].$$
\end{definition}

To illustrate that our definitions of SID are necessary and appropriate, we provide definitions of integrated dispersion in Section {\secintegrated} of the supplementary material, followed by three examples.

\section{Probabilistic construction of IOD}
\label{sec:Overdispersion}


In this section, we aim to generate a new model $\X$ having SIOD based on a GDM $\Z$ having SIED. We consider $\Z$ in a general form in the way that conditional on $\Z(t)=\z$,
each flow $N^{\Z}_{vv'}$ over arrow $(v,v')\in A$ is associated with a general TRF $\Upsilon_{vv'}$ which depends on time $t$ and state of the graph $\z$, such that 
\begin{equation*}
	Q(t,(z_v,z_{v'},\z_{V\backslash \{v,v'\}}),(z_v-1,z_{v'}+1, \z_{V\backslash \{v,v'\}}))=\Upsilon_{vv'}(t,\z).
\end{equation*}
The Markov chain interpretation of $\Z$ can be specified by
the infinitesimal transition probabilities:
\begin{equation}
	\begin{split}
		\label{eqn:infinitesimal_transition_probabilities}
		&\mathbb{P}(\Delta^{\Z}_{vv'}(t,h)=0\mid \Z(t)=\z)=1-\Upsilon_{vv'}(t,\z)h+o(h),\\
		&\mathbb{P}(\Delta^{\Z}_{vv'}(t,h)=1\mid \Z(t)=\z)=\Upsilon_{vv'}(t,\z)h+o(h),\\
		&\mathbb{P}(\Delta^{\Z}_{vv'}(t,h)>1\mid \Z(t)=\z)=o(h),\\
		&\mathbb{P}(\Delta^{\Z}_{vv'}(t,h)<0\mid \Z(t)=\z)=0.
	\end{split}
\end{equation}
Without loss of generality, we suppose the initial values of the dynamics over the graph are integers for notational simplicity.

\subsection{IOD construction with boundedness constraints}
\label{sec:iod_bounded}
In this subsection, we focus on generating GDMs having SIOD over outgoing arrows with the same tail.
We consider the case that there are multiple connected outgoing arrows of vertex $v$ such that $|N_G^{+}(v)|\geq 1$, where $N_G^{+}(v)$ is the set of vertices $v'\in V$ such that  $(v, v')\in A$ and $|N_G^{+}(v)|$ is its cardinality.  
Suppose $N_G^{+}(v)=\{v_1',\ldots,v_m'\}$ and $|N_G^{+}(v)|=m$.
The transition rate of $N^{\Z}_{vv_i'}(t)$ for $i\in \{1,\ldots,m\}$ is given by
\begin{equation*}
	Q(t,(z_v,z_{v_i'},\z_{V\backslash \{v,v_i'\}}),(z_v-1,z_{v_i'}+1, \z_{V\backslash \{v,v_i'\}}))=r_{vv_i'}(t,\z)z_v\mathbbm{1}_{\{z_v\geq 1\}},
\end{equation*}
where $\z=(z_v,z_{v_i'},\z_{V\backslash \{v,v_i'\}})$.
Considering a sufficiently short period $[t,t+h]$, by \eqref{eqn:infinitesimal_transition_probabilities} the probability that one transition from vertex $v$ to vertex $v_i'$ for $i\in \{1,\ldots,m\}$ is given by
$$\mathbb{P}(\Delta^{\Z}_{vv_i'}(t,h)=1\mid \Z(t)=\z)=r_{vv_i'}(t,\z)z_vh+o(h).$$
For notational convenience, denote $\Delta^{\Z}_{vv_0'}(t,h)$ as the remaining individuals at vertex $v$. 
Then the joint distribution of $\{\Delta^{\Z}_{vv_i'}(t,h)=k_{i}\}_{i\in \{0,\ldots,m\}}$ is given by
\begin{align}
	\label{eqn:multinomial_pdf}
	\mathbb{P}(\{\Delta^{\Z}_{vv_i'}(t,h)=k_{i}\}_{i\in \{0,\ldots,m\}} \mid \Z(t)=\z)=\frac{\Gamma(z_v+1)}{\prod_{i=0}^{m}\Gamma(k_{i}+1)}\prod_{i=0}^{m}\left[\widetilde{\pi}_{vv_i'}(t,h,\z)\right]^{k_{i}}+o(h),
\end{align}
where $\Gamma(\cdot)$ is the gamma function, $z_v\geq 1$ and $k_{i}\in \{0,1,\ldots,z_v\}$ for $i\in \{0,\ldots,m\}$ such that $\sum_{i=0}^{m}k_{i}=z_v$.  
Here, for $i\in \{1,\ldots,m\}$
\begin{align}
	\label{eqn:widetilde_pi}
	\widetilde{\pi}_{vv_i'}(t,h,\z)\nonumber&=\left(1-e^{-\sum_{j=1}^{m}\int_{t}^{t+h}r_{vv_j'}(s,\z)ds}\right)\frac{r_{vv_i'}(t,\z)z_vh+o(h)}{\sum_{j=1}^{m}r_{vv_j'}(t,\z)z_vh+o(h)}+o(h)\nonumber\\
	&=\left(1-e^{-\sum_{j=1}^{m}\int_{t}^{t+h}r_{vv_j'}(s,\z)ds}\right)\frac{r_{vv_i'}(t,\z)z_vh}{\sum_{j=1}^{m}r_{vv_j'}(t,\z)z_vh+o(h)}+o(h)\nonumber\\
	&=\left(1-e^{-\sum_{j=1}^{m}\int_{t}^{t+h}r_{vv_j'}(s,\z)ds}\right)\frac{r_{vv_i'}(t,\z)z_vh}{\sum_{j=1}^{m}r_{vv_j'}(t,\z)z_vh}\left( \frac{1}{1+o(h)} \right)+o(h)\nonumber\\
	&=\left(1-e^{-\sum_{j=1}^{m}\int_{t}^{t+h}r_{vv_j'}(s,\z)ds}\right)\frac{r_{vv_i'}(t,\z)}{\sum_{j=1}^{m}r_{vv_j'}(t,\z)}\left(1+o(h) \right)+o(h)\nonumber\\
	&=\left(1-e^{-\sum_{j=1}^{m}\int_{t}^{t+h}r_{vv_j'}(s,\z)ds}\right)\frac{r_{vv_i'}(t,\z)}{\sum_{j=1}^{m}r_{vv_j'}(t,\z)}+o(h),
\end{align}
where we used Taylor series in the fourth equality,
and $$\widetilde{\pi}_{vv_0'}(t,h,\z)=1-\sum_{i=1}^{m}\widetilde{\pi}_{vv_i'}(t,h,\z).$$ 
Plugging \eqref{eqn:widetilde_pi} into \eqref{eqn:multinomial_pdf}, we can rewrite \eqref{eqn:multinomial_pdf} as
\begin{align}
	\label{eqn:multinomial_pdf_simplified2}
	\mathbb{P}(\{\Delta^{\Z}_{vv_i'}(t,h)=k_{i}\}_{i\in \{0,\ldots,m\}} \mid \Z(t)=\z)=\frac{\Gamma(z_v+1)}{\prod_{i=0}^{m}\Gamma(k_{i}+1)}\prod_{i=0}^{m}\left[\pi_{vv_i'}(t,h,\z)\right]^{k_{i}}+o(h),
\end{align}
where 
\begin{align*}
	\pi_{vv_i'}(t,h,\z)=\left\{ \begin{array}{ll} 
		\left(1-e^{-\sum_{j=1}^{m}\int_{t}^{t+h}r_{vv_j'}(s,\z)ds}\right)\frac{r_{vv_i'}(t,\z)}{\sum_{j=1}^{m}r_{vv_j'}(t,\z)} &i\in \{1,\ldots,m\}, \\
		1-\sum_{j=1}^{m}\pi_{vv_j'}(t,h,\z) & i=0.
	\end{array}\right.
\end{align*}

The following proposition shows that a GDM $\X$ having SIOD can be generated over connected outgoing arrows $\{(v,v_i)\}_{i\in \{1,\ldots,m\}}$.
\begin{proposition}
	\label{prop:iod_multinomial}
	Suppose that $r_{vv_i'}(t,\x)$, for each $i\in \{1,\ldots,m\}$, is a positive function that is uniformly
	continuous in $t$. Further suppose that $\{\Delta^{\X}_{vv_i'}(t,h)=k_{i}\}_{i\in \{0,\ldots,m\}}$ are jointly distributed in a sufficiently short period $[t,t+h]$ as follows: 
	\begin{align}
		\label{eqn:multinomial_pdf_modified}
		&\hspace*{-0.5cm}\mathbb{P}(\{\Delta^{\X}_{vv_i'}(t,h)=k_{i}\}_{i\in \{0,\ldots,m\}} \mid \X(t)=\x,\{\Pi_{vv_i'}(t,h,\x)\}_{i\in \{0,\ldots,m\}})\nonumber\\
		&=\frac{\Gamma(x_v+1)}{\prod_{i=0}^{m}\Gamma(k_{i}+1)}\prod_{i=0}^{m}\left(\Pi_{vv_i'}(t,h,\x)\right)^{k_{i}}+o(h),
	\end{align}
	where $x_v\geq 1$ and $k_{i}\in \{0,1,\ldots,x_v\}$ for $i\in \{0,\ldots,m\}$ such that $\sum_{i=0}^{m}k_{i}=x_v$. 
	Further suppose that the family $\{\Pi_{vv_i'}(t,h,\x)\}_{i\in \{0,1,\ldots,m\}}$ is distributed according to the Dirichlet distribution $\operatorname{Dir}(\{\alpha_{vv_i'}(t,h,\x)\}_{i\in \{0,1,\ldots,m\}})$ having $$\alpha_{vv_i'}(t,h,\x)=c\pi_{vv_i'}(t,h,\x) \quad\quad\text{for}\; i \in \{0,\ldots,m\},$$
	where $c>0$ is an 
	inverse noise parameter and 
	\begin{align*}
		\pi_{vv_i'}(t,h,\x)=\left\{ \begin{array}{ll} 
			\left(1-e^{-\sum_{j=1}^{m}\int_{t}^{t+h}r_{vv_j'}(s,\x)ds}\right)\frac{r_{vv_i'}(t,\x)}{\sum_{j=1}^{m}r_{vv_j'}(t,\x)} &i\in \{1,\ldots,m\}, \\
			1-\sum_{j=1}^{m}\pi_{vv_j'}(t,h,\x) & i=0.
		\end{array}\right.
	\end{align*}
	The following results hold:
	\begin{enumerate}
		\item For each $i\in \{1,\ldots,m\}$, the infinitesimal mean $\mu_{vv_i}^{d\X}(t,\x)$ is given by 
		$$\mu_{vv_i'}^{d\X}(t,\x)=x_v r_{vv_i'}(t,\x)$$
		and the infinitesimal variance $[\sigma_{vv_i}^{d\X}(t,\x)]^2$ is given by
		$$[\sigma_{vv_i'}^{d\X}(t,\x)]^2=(1+(x_v-1)(c+1)^{-1})x_v r_{vv_i'}(t,\x).$$
		When $x_v>1$, $\X(t)$  has IOD at $\X(t)=\x$ with respect to each arrow of $\{(v,v_i')\}_{i\in \{1,\ldots,m\}}$ and $\X(t)$ has SIOD at $\X(t)=\x$ for  connected outgoing arrows $\{(v,v_i')\}_{i\in \{1,\ldots,m\}}$; when $x_v=1$, $\X(t)$ has SIED at $\X(t)=\x$ for connected outgoing arrows $\{(v,v_i')\}_{i\in \{1,\ldots,m\}}$.
		Furthermore, for $i,j\in \{1,\ldots,m\}$ and $i\neq j$, the infinitesimal covariance $\sigma_{vv_i',vv_j'}^{d\X}(t,\x)=0.$
		
		\smallskip
		\item 	Denote $\mathcal{S}$ as the set of transitions over arrows $\{(v,v_i)\}_{i\in \{1,\ldots,m\}}$, i.e.,
		\begin{align}
			\label{eqn:setA_multinomial}
			\mathcal{S}:=\left\{k_i\geq 1 \;\text{for } i\in \{1,\ldots,m\};\; \sum_{i=0}^m k_i=x_v\right\}.
		\end{align}
		Then the conditional probability that transitions happen over two or more arrows
		\begin{align*}
			\mathbb{P}(\{\Delta^{\X}_{vv_i'}(t,h)=k_{i}\}_{i\in \{0,\ldots,m\}},\; |\mathcal{S}|\geq 2  \mid \X(t)=\x)=o(h)
		\end{align*}
		and the conditional probability that only one transition happens over a single arrow
		\begin{align}
			\label{eqn:transition_multinormial_00}
			\mathbb{P}(\{\Delta^{\X}_{vv_i'}(t,h)=k_{i}\}_{i\in \{0,\ldots,m\}},\; |\mathcal{S}|=1  \mid \X(t)=\x)=\sum_{i=1}^{m}q_{{v}{v_i'}}(t,\x,k_i)h+o(h),
		\end{align}
		where $|\mathcal{S}|$ is the cardinality of $\mathcal{S}$ and for $i\in \{1,\ldots,m\}$
		\begin{align}
			\label{eqn:rate00}
			q_{{v}{v_i'}}(t,\x,k_i)
			=c{x_v \choose k_i}\frac{\Gamma(k_i)\Gamma(x_v-k_i+c)}{\Gamma(x_v+c)}r_{vv_i'}(t,\x).
		\end{align}
	\end{enumerate}
\end{proposition}

The proof of Proposition \ref{prop:iod_multinomial} is postponed to Section \ref{Section:proof_multinomial} in the supplemental material.
We note that a crucial difference between equations \eqref{eqn:infinitesimal_transition_probabilities}
and \eqref{eqn:transition_multinormial_00}
is that, one transition over any single arrow has up to $m(\geq 1)$ units in \eqref{eqn:transition_multinormial_00} while one transition over a specific arrow has exactly one unit in \eqref{eqn:infinitesimal_transition_probabilities}. 
From Proposition \ref{prop:iod_multinomial}, we can see that the methodology proposed in \cite{breto2011compound} is even a special case of our $m=1$ case. Now, we realize the methodology proposed in Proposition \ref{prop:iod_multinomial} in the Algorithm \ref{fig:euler_beta}.
\begin{algorithm}[!htb]
	\noindent\begin{tabular}{l}
		Set the initial value $\X(0)$ and time interval $[0,T]$.\\
		Set time increment $\delta=T/N$ for integer $N$; define $t_n=n\delta$.\\
		FOR $n=0 \mbox{ to } N-1$\\
		\asp     FOR each $v \in V$ with $N_G^{+}(v)=\{v_1',\ldots,v_m'\}$ \mystretch\\
		\asp \asp Generate $\{\Pi_{vv_i'}\}_{i\in \{0,1,\ldots,m\}}$  according to the Dirichlet distribution  \mystretch\\
		\asp \asp\asp $\operatorname{Dir}(\{\alpha_{vv_i'}\}_{i\in \{0,1,\ldots,m\}})$ having \mystretch\\
		\asp \asp\asp $\alpha_{vv_i'}=c\pi_{vv_i'} \quad\text{for}\; i \in \{1,\ldots,m\}\quad\text{and}\quad \alpha_{vv_0'}=c-\sum_{i=1}^{m}\alpha_{vv_i'},$\mystretch\\
		\asp \asp\asp  where for $i\in \{1,\ldots,m\}$\mystretch\\
		\asp \asp\asp $\pi_{vv_i'}=\left(1-e^{-\sum_{j=1}^{m}\int_{t_n}^{t_{n+1}}r_{vv_j'}(s,\x)ds}\right)\frac{r_{vv_i'}(t_n,\x)}{\sum_{j=1}^{m}r_{vv_j'}(t_n,\x)}$.	\mystretch\\	
		\asp \asp  Generate process increments \mystretch\\
		\asp \asp\asp  $\{\Delta^{\X}_{vv_i'}\}_{i\in \{0,1,\ldots,m\}}\sim \mbox{Multinomial}(X_v(t_n),\{\Pi_{vv_i'}\}_{i\in \{0,1,\ldots,m\}})$\mystretch\\
		\asp \asp\asp  where $\Delta^{\X}_{vv_0'}$ stands for retain individuals.\mystretch\\
		\asp\asp Set $X_v(t_{n+1})=\Delta^{\X}_{vv_0'}+\sum_{\overline{v}\in N_G^{-}(v)}\Delta^{\X}_{\overline{v}v}$\mystretch\\
		\asp END FOR\mystretch\\
		END FOR
	\end{tabular}
	\caption{Euler scheme on generating dynamics having IOD over connected outgoing arrows $\{(v,v_i)\}_{i\in \{1,\ldots,m\}}$.}
	\label{fig:euler_beta}
\end{algorithm}

\subsection{IOD construction without boundedness constraints}
\label{sec:iod_unbounded}
Unbounded processes such as the pure birth process, have wide applications. In this subsection, we focus on generating GDMs having IOD over incoming arrows with the same head without boundedness constraints.

We consider the case that there are multiple connected incoming arrows of vertex $u'$ such that $|N_G^{-}(u')|\geq 1$, where $N_G^{-}(u')$ is the set of vertices $u\in V$ such that  $(u, u')\in A$ and $|N_G^{-}(u')|$ is its cardinality.  
Suppose $N_G^{-}(u'):=\{u_1,\ldots,u_{\overline{m}}\}$ and $\overline{m}:=|N_G^{-}(u')|$. 
The transition probability of $N^{\Z}_{u_i u'}(t)$ for $i\in \{1,\ldots,\overline{m}\}$ is given by
\begin{align*}
	Q(t,(z_{u_i},z_{u'},\z_{V\backslash \{u_i,u'\}}),(z_{u_i}-1,z_{u'}+1, \z_{V\backslash \{u_i,u'\}}))=r_{u_i u'}(t,\z)z_{u'}\mathbbm{1}_{\{z_{u'}>0\}},
\end{align*}
where $\z=(z_{u_i},z_{u'},\z_{V\backslash \{u_i,u'\}})$.
Considering a sufficiently short period $[t,t+h]$, by \eqref{eqn:infinitesimal_transition_probabilities} the probability that one transition from vertex $u_i$ to vertex $u'$
$$\mathbb{P}(\Delta^{\Z}_{u_i u'}(t,h)=1\mid \Z(t)=\z)=r_{u_i u'}(t,\z)z_{u'} h+o(h).$$
The joint distribution of increments of $\{N^{\Z}_{u_i u'}(t)\}_{i\in \{1,\ldots,\overline{m}\}}$ is given by
\begin{align}
	\label{eqn:negative_multinomial_pdf}
	&\mathbb{P}(\{\Delta^{\Z}_{u_i u'}(t,h)=k_{i}\}_{i\in \{1,\ldots,\overline{m}\}} \mid \Z(t)=\z)\\
	&=\frac{\Gamma(z_{u'}+\sum_{i=1}^{\overline{m}}k_{i})}{\Gamma(z_{u'})\prod_{i=1}^{\overline{m}}\Gamma(k_{i}+1)}\left[1-\sum_{i=1}^{\overline{m}}\widetilde{\pi}_{u_i u'}(t,h,\z)\right]^{z_{u'}}\prod_{i=0}^{\overline{m}}\left[\widetilde{\pi}_{u_i u'}(t,h,\z)\right]^{k_{i}}+o(h),\nonumber
\end{align}
where $z_{u'}>0$ and $k_{i}\in \{0,1,2,\ldots\}$ for $i\in \{1,\ldots,\overline{m}\}$.
Here, for $i\in \{1,\cdots,\overline{m}\}$
\begin{align}
	\label{eqn:widetilde_pi_negative_multinomial}
	\widetilde{\pi}_{u_i u'}(t,h,\z)
	&=\left(1-e^{-\sum_{j=1}^{\overline{m}}\int_{t}^{t+h}r_{u_i u'}(s,\z)ds}\right)\frac{r_{vv_i'}(t,\z)z_{u'}h+o(h)}{\sum_{j=1}^{\overline{m}}r_{u_i u'}(t,\z)z_{u'}h+o(h)}+o(h)\nonumber\\
	&=\left(1-e^{-\sum_{j=1}^{\overline{m}}\int_{t}^{t+h}r_{u_i u'}(s,\z)ds}\right)\frac{r_{vv_i'}(t,\z)z_{u'}h}{\sum_{j=1}^{\overline{m}}r_{u_i u'}(t,\z)z_{u'}h+o(h)}+o(h)\nonumber\\
	&=\left(1-e^{-\sum_{j=1}^{\overline{m}}\int_{t}^{t+h}r_{u_i u'}(s,\z)ds}\right)\frac{r_{vv_i'}(t,\z)z_{u'}h}{\sum_{j=1}^{\overline{m}}r_{u_i u'}(t,\z)z_{u'}h}\left( \frac{1}{1+o(h)} \right)+o(h)\nonumber\\
	&=\left(1-e^{-\sum_{j=1}^{\overline{m}}\int_{t}^{t+h}r_{u_i u'}(s,\z)ds}\right)\frac{r_{vv_i'}(t,\z)}{\sum_{j=1}^{\overline{m}}r_{u_i u'}(t,\z)}\left(1+o(h) \right)+o(h)\nonumber\\
	&=\left(1-e^{-\sum_{j=1}^{\overline{m}}\int_{t}^{t+h}r_{u_j u'}(s,\z)ds}\right)\frac{r_{u_i u'}(t,\z)}{\sum_{j=1}^{\overline{m}}r_{u_j u'}(t,\z)}+o(h),
\end{align} 
where we used Taylor series in the fourth equality. 
Plugging \eqref{eqn:widetilde_pi_negative_multinomial} into \eqref{eqn:negative_multinomial_pdf}, we can rewrite \eqref{eqn:negative_multinomial_pdf} as
\begin{align*}
	&\mathbb{P}(\{\Delta^{\Z}_{u_i u'}(t,h)=k_{i}\}_{i\in \{1,\ldots,\overline{m}\}} \mid \Z(t)=\z)\\
	&=\frac{\Gamma(z_{u'}+\sum_{i=1}^{\overline{m}}k_{i})}{\Gamma(z_{u'})\prod_{i=1}^{\overline{m}}\Gamma(k_{i}+1)}\left[\pi_{u_0 u'}(t,h,\z)\right]^{z_{u'}}\prod_{i=0}^{\overline{m}}\left[\pi_{u_i u'}(t,h,\z)\right]^{k_{i}}+o(h),\nonumber
\end{align*}
where 
\begin{align*}
	\pi_{u_i u'}(t,h,\z)=\left\{ \begin{array}{ll} 
		\left(1-e^{-\sum_{j=1}^{\overline{m}}\int_{t}^{t+h}r_{u_j u'}(s,\z)ds}\right)\frac{r_{u_i u'}(t,\z)}{\sum_{j=1}^{\overline{m}}r_{u_j u'}(t,\z)} &i\in \{1,\ldots,\overline{m}\}, \\
		1-\sum_{j=1}^{\overline{m}}\pi_{u_j u'}(t,h,\z) & i=0.
	\end{array}\right.
\end{align*}


The following proposition shows that a GDM $\X$ having SIOD can be generated over connected incoming arrows $\{(u_i, u')\}_{i\in \{1,\ldots,\overline{m}\}}$.
\begin{proposition}
	\label{prop:iod_negative_multinomial}
	Suppose that $r_{u_i u'}(t,\x)$, for each $i\in \{1,\ldots,\overline{m}\}$, is a positive function that is uniformly
	continuous in $t$. Further suppose that  the increments of $\{N^{\X}_{u_i u'}(t)\}_{i\in \{1,\ldots,\overline{m}\}}$ are jointly distributed in a sufficiently short period $[t,t+h]$ as: 
	\begin{align}
		\label{eqn:negative_multinomial_pdf_modified}
		&\mathbb{P}(\{\Delta^{\X}_{u_i u'}(t,h)=k_{i}\}_{i\in \{1,\ldots,\overline{m}\}} \mid \X(t)=\x,\{\Pi_{u_i u'}(t,h,\x)\}_{i\in \{0,1,\ldots,\overline{m}\}})\nonumber\\
		&=\frac{\Gamma(x_{u'}+\sum_{i=1}^{\overline{m}}k_{i})}{\Gamma(x_{u'})\prod_{i=1}^{\overline{m}}\Gamma(k_{i}+1)}\left[\Pi_{u_0 u'}(t,h,\x)\right]^{x_{u'}}\prod_{i=1}^{\overline{m}}\left[\Pi_{u_i u'}(t,h,\x)\right]^{k_{i}}+o(h),
	\end{align}
	where $x_{u'}>0$ and $k_{i}\in \{0,1,2,\ldots\}$ for $i\in \{1,\ldots,\overline{m}\}$. Here, $\{\Pi_{u_i u'}(t,h,\x)\}_{i\in \{0,1,\ldots,\overline{m}\}}$ is distributed according to the Dirichlet distribution $\operatorname{Dir}(\{\alpha_{u_i u'}(t,h,\x)\}_{i\in \{0,1,\ldots,\overline{m}\}})$ having $$\alpha_{u_i u'}(t,h,\x)=c\pi_{u_i u'}(t,h,\x) \quad\quad\text{for}\; i \in \{0,1,\ldots,\overline{m}\} $$
	where $c>0$ is an 
	inverse noise parameter and 
	\begin{align*}
		\pi_{u_i u'}(t,h,\x)=\left\{ \begin{array}{ll} 
			\left(1-e^{-\sum_{j=1}^{\overline{m}}\int_{t}^{t+h}r_{u_j u'}(s,\x)ds}\right)\frac{r_{u_i u'}(t,\x)}{\sum_{j=1}^{\overline{m}}r_{u_j u'}(t,\x)} &i\in \{1,\ldots,\overline{m}\}, \\
			1-\sum_{j=1}^{\overline{m}}\pi_{u_j u'}(t,h,\x) & i=0.
		\end{array}\right.
	\end{align*} 
	The following results hold:
	\begin{enumerate}
		\item When $c>2e^{\sum_{i=1}^{\overline{m}}\int_{t}^{t+h}r_{u_iu'}(s,\x)ds}$, for any $i\in \{1,\ldots,\overline{m}\}$, the infinitesimal mean $\mu_{u_i u'}^{d\X}(t,\x)$ is given by 		
		$$\mu_{u_i u'}^{d\X}(t,\x)=x_{u'} r_{u_iu'}(t,\x)\frac{c}{c-1},$$
		and the infinitesimal variance $[\sigma_{u_i u'}^{d\X}(t,\x)]^2$ is given by
		$$[\sigma_{u_i u'}^{d\X}(t,\x)]^2=x_{u'}^2 r_{u_iu'}(t,\x) \frac{c}{(c-1)(c-2)}+x_{u'} r_{u_iu'}(t,\x) \frac{c}{c-2}.$$		
		Then $\X(t)$ has IOD at $\X(t)=\x$ with respect to each arrow of $\{(u_i, u')\}_{i\in \{1,\ldots,\overline{m}\}}$, and $\X(t)$ has SIOD at $\X(t)=\x$ for connected incoming arrows $\{(u_i, u')\}_{i\in \{1,\ldots,\overline{m}\}}$.
		Furthermore, when $c>2e^{\sum_{i=1}^{\overline{m}}\int_{t}^{t+h}r_{u_iu'}(s,\x)ds}$, for $i,j\in \{1,\ldots,\overline{m}\}$ and $i\neq j$, the infinitesimal covariance $\sigma_{uu_i',uu_j'}^{d\X}(t,\x)=0.$			\smallskip
		
		\item 	Denote $\mathcal{S}$ as the set of transitions over arrows $\{(u_i, u')\}_{i\in \{1,\ldots,\overline{m}\}}$, i.e.,
		\begin{align}
			\label{eqn:setA_negative_multinomial}
			\overline{\mathcal{S}}:=\left\{k_i\geq 1 \;\text{for } i\in \{1,\ldots,\overline{m}\}\right\}.
		\end{align}
		Then the conditional probability that transitions happen over two or more arrows
		\begin{align*}
			\mathbb{P}(\{\Delta^{\X}_{u_i u'}(t,h)=k_{i}\}_{i\in \{1,\ldots,\overline{m}\}},\; |\overline{\mathcal{S}}|\geq 2  \mid \X(t)=\x)=o(h),
		\end{align*}
		and the conditional probability that only one transition happens over a single arrow
		\begin{align*}
			\mathbb{P}(\{\Delta^{\X}_{u_i u'}(t,h)=k_{i}\}_{i\in \{1,\ldots,\overline{m}\}},\; |\overline{\mathcal{S}}|=1  \mid \X(t)=\x)=\sum_{i=1}^{\overline{m}}q_{u_i u'}(t,\x,k_i)h+o(h),
		\end{align*}
		where $|\overline{\mathcal{S}}|$ is the cardinality of $\overline{\mathcal{S}}$ and for $i\in \{1,\ldots,\overline{m}\}$
		\begin{align}
			\label{eqn:rate02}
			q_{u_i u'}(t,\x,k_i)=c\frac{\Gamma(x_{u'}+\sum_{i=1}^{\overline{m}}k_{i})}{\Gamma(x_{u'})\prod_{i=1}^{\overline{m}}\Gamma(k_{i}+1)}\frac{\Gamma(x_{u'}+c)\Gamma(k_{i})}{\Gamma(x_{u'}+\sum_{i=1}^{\overline{m}}k_{i}+c)}r_{u_iu'}(t,\x).
		\end{align}
	\end{enumerate}
\end{proposition}

The proof of Proposition \ref{prop:iod_negative_multinomial} is postponed to Section \ref{Section:proof_negative_multinomial} in the supplemental material.
We note that in practice, users of Proposition \ref{prop:iod_negative_multinomial} can simply treat $c$ as an unknown parameter through using modern likelihood-based parameter inference algorithms such as \cite{ionides2015inference}, as illustrated in Section \ref{sec:application}. 
\section*{Acknowledgments}
This research project was partially supported by NSF grant DMS-$1761603$. Ning Ning's research was also partially supported by the Seed Fund Grant Award at Texas A\&M University. 

\begin{table}[h!]
	\begin{center}
		\caption{Notation.}	\label{tab:TableOfNotation}
		\begin{tabular}{r c p{12cm} }
			\toprule
			$G = (V, A)$ && $G$: graph, $V$: set of vertices, and $A$: set of arrows, Sect. \ref{sec:Model_Setup_graph}.\\
			$S_o$ && Set of all source vertices, Sect. \ref{sec:Model_Setup_graph}.\\
			$S_i$ && Set of all sink vertices, Sect. \ref{sec:Model_Setup_graph}.\\
			$X_v(t)$ && Stochastic process on vertex $v\in V$ at time $t\in [0,\infty)$, Sect. \ref{sec:Model_Setup_graph}.\\
			$N^{\X}_{vv'}(t)$ && Nondecreasing integer-valued jump process over $(v, v')$, Sect. \ref{sec:Model_Setup_graph}.\\
			$q(t,\x,\mathbf{l})$ && TRF, Sect. \ref{sec:Model_Setup_dynamics}.\\
			$q_{\{{v_i'}{v_i''}\}_{i\in |A|}}(\cdot,\cdot,\cdot)$ && TRF, Sect. \ref{sec:Model_Setup_dynamics}.\\
			$Q(t,\x,\x')$ && TRF, Eqn.  \eqref{eqn:transition_rate_Q}.\\
			$\Delta^{\X}_{vv'}(t,h)$ && Increments of $N^{\X}_{vv'}(t)$ in time interval $[t,t+h]$, Sect. \ref{sec:Model_Setup_dispersion}.\\
			$[\sigma_{vv'}^{d\X}(t,\x)]^2$ && Infinitesimal variance, Def. \ref{def:infinitesimal_dispersion}.\\
			$\mu_{vv'}^{d\X}(t,\x)$ && Infinitesimal mean, Def. \ref{def:infinitesimal_dispersion}.\\
			$D_{vv'}^{d\X}(t,\x)$ && Infinitesimal dispersion index, Def. \ref{def:infinitesimal_dispersion}.\\
			$\Upsilon_{vv'}(t,\z)$ && Transition probability over $(v,v')$, Sect. \ref{sec:Overdispersion}.\\	
			$r_{\cdot \cdot}(t,\z)$ && Per-capita rate function, Sect. \ref{sec:Overdispersion}.\\
			$\widetilde{\pi}_{\cdot \cdot}(t,h,\z)$&& Transition probability, Sect. \ref{sec:Overdispersion}.\\
			$\pi_{\cdot \cdot}(t,h,\z)$&& Transition probability after rewriting, Sect. \ref{sec:Overdispersion}.\\
			$\Pi_{\cdot \cdot}(t,h,\x)$ && Stochastic transition probability, Sect. \ref{sec:Overdispersion}.\\
			$c>0$  &&  Inverse noise parameter, Sect. \ref{sec:Overdispersion}.\\
			$\alpha_{\cdot \cdot}(t,h,\x)$ && Parameter in the distribution of $\Pi_{\cdot \cdot}(t,h,\x)$, Sect. \ref{sec:Overdispersion}.\\
			$\beta_{\cdot \cdot}(t,h,\x)$ && Parameter in the distribution of $\Pi_{\cdot \cdot}(t,h,\x)$, Sect. \ref{sec:Overdispersion}.\\
			$\mathcal{S}$ && Set defined in Eqn. \eqref{eqn:setA_multinomial}.\\			
			$\overline{\mathcal{S}}$ && Set defined in Eqn. \eqref{eqn:setA_negative_multinomial}.\\
			$[\sigma_{vv'}^{\X}(t,\x_0)]^2$ && Integrated variance, Def. \ref{def:integrated_dispersion}.\\
			$\mu_{vv'}^{\X}(t,\x_0)$ && Integrated mean, Def. \ref{def:integrated_dispersion}.\\
			$D_{vv'}^{\X}(t,\x_0)$ && Integrated dispersion index, Def. \ref{def:integrated_dispersion}.\\
			\bottomrule
		\end{tabular}
	\end{center}	
\end{table}



\bibliography{bib-ms}

\begin{thebibliography}{}

\bibitem[Albrecher et~al., 2017]{albrecher2017reinsurance}
Albrecher, H., Beirlant, J., and Teugels, J.~L. (2017).
\newblock {\em Reinsurance: {A}ctuarial and statistical aspects}.
\newblock John Wiley \& Sons.

\bibitem[Berk and MacDonald, 2008]{berk2008overdispersion}
Berk, R. and MacDonald, J.~M. (2008).
\newblock Overdispersion and {P}oisson regression.
\newblock {\em Journal of Quantitative Criminology}, 24(3):269--284.

\bibitem[Bj{\o}rnstad et~al., 2002]{bjornstad2002dynamics}
Bj{\o}rnstad, O.~N., Finkenst{\"a}dt, B.~F., and Grenfell, B.~T. (2002).
\newblock Dynamics of measles epidemics: {E}stimating scaling of transmission
  rates using a time series {SIR} model.
\newblock {\em Ecological monographs}, 72(2):169--184.

\bibitem[Bj{\o}rnstad and Grenfell, 2001]{bjornstad2001noisy}
Bj{\o}rnstad, O.~N. and Grenfell, B.~T. (2001).
\newblock Noisy clockwork: {T}ime series analysis of population fluctuations in
  animals.
\newblock {\em Science}, 293(5530):638--643.

\bibitem[Bret{\'o}, 2012]{breto2012infinitesimal}
Bret{\'o}, C. (2012).
\newblock On the infinitesimal dispersion of multivariate {M}arkov counting
  systems.
\newblock {\em Statistics \& Probability Letters}, 82(4):720--725.

\bibitem[Bret{\'o}, 2021]{breto2021co}
Bret{\'o}, C. (2021).
\newblock Co-jumps and {M}arkov counting systems in random environments.
\newblock In {\em Contemporary Approaches and Methods in Fundamental
  Mathematics and Mechanics}, pages 277--292. Springer.

\bibitem[Bret{\'o} et~al., 2009]{breto2009time}
Bret{\'o}, C., He, D., Ionides, E.~L., and King, A.~A. (2009).
\newblock Time series analysis via mechanistic models.
\newblock {\em The Annals of Applied Statistics}, 3(1):319--348.

\bibitem[Bret{\'o} and Ionides, 2011]{breto2011compound}
Bret{\'o}, C. and Ionides, E.~L. (2011).
\newblock Compound {M}arkov counting processes and their applications to
  modeling infinitesimally over-dispersed systems.
\newblock {\em Stochastic Processes and their Applications},
  121(11):2571--2591.

\bibitem[Brown et~al., 1998]{brown1998variance}
Brown, T.~C., Hamza, K., and Xia, A. (1998).
\newblock On the variance to mean ratio for random variables from {M}arkov
  chains and point processes.
\newblock {\em Journal of applied probability}, pages 303--312.

\bibitem[Chen et~al., 2017]{chen2017network}
Chen, S., Shojaie, A., and Witten, D.~M. (2017).
\newblock Network reconstruction from high-dimensional ordinary differential
  equations.
\newblock {\em Journal of the American Statistical Association},
  112(520):1697--1707.

\bibitem[Ding and Ning, 2021]{ding2021markov}
Ding, K. and Ning, N. (2021).
\newblock Markov chain approximation and measure change for time-inhomogeneous
  stochastic processes.
\newblock {\em Applied Mathematics and Computation}, 392:125732.

\bibitem[Gao and Sendova, 2022]{gao2022applications}
Gao, D. and Sendova, K.~P. (2022).
\newblock Applications of the classical compound {P}oisson model with claim
  sizes following a compound distribution.
\newblock {\em Probability in the Engineering and Informational Sciences},
  pages 1--30.

\bibitem[Gillespie, 1984]{gillespie1984molecular}
Gillespie, J.~H. (1984).
\newblock The molecular clock may be an episodic clock.
\newblock {\em Proceedings of the National Academy of Sciences},
  81(24):8009--8013.

\bibitem[He et~al., 2010]{he2010plug}
He, D., Ionides, E.~L., and King, A.~A. (2010).
\newblock Plug-and-play inference for disease dynamics: measles in large and
  small populations as a case study.
\newblock {\em Journal of the Royal Society Interface}, 7(43):271--283.

\bibitem[Ionides et~al., 2021]{ionides2020bagged}
Ionides, E.~L., Asfaw, K., Park, J., and King, A.~A. (2021).
\newblock Bagged filters for partially observed spatiotemporal systems.
\newblock {\em Journal of the American Statistical Association, DOI:
  10.1080/01621459.2021.1974867}.

\bibitem[Ionides et~al., 2011]{ionides2011iterated}
Ionides, E.~L., Bhadra, A., Atchad{\'e}, Y., and King, A. (2011).
\newblock Iterated filtering.
\newblock {\em The Annals of Statistics}, 39(3):1776--1802.

\bibitem[Ionides et~al., 2006]{ionides2006inference}
Ionides, E.~L., Bret{\'o}, C., and King, A.~A. (2006).
\newblock Inference for nonlinear dynamical systems.
\newblock {\em Proceedings of the National Academy of Sciences},
  103(49):18438--18443.

\bibitem[Ionides et~al., 2015]{ionides2015inference}
Ionides, E.~L., Nguyen, D., Atchad{\'e}, Y., Stoev, S., and King, A.~A. (2015).
\newblock Inference for dynamic and latent variable models via iterated,
  perturbed {B}ayes maps.
\newblock {\em Proceedings of the National Academy of Sciences},
  112(3):719--724.

\bibitem[Jayaraman and Matis, 2010]{jayaraman2010batch}
Jayaraman, R. and Matis, T.~I. (2010).
\newblock Batch arrivals and service—single station queues.
\newblock {\em Wiley Encyclopedia of Operations Research and Management
  Science}.

\bibitem[Katzfuss et~al., 2020]{katzfuss2020ensemble}
Katzfuss, M., Stroud, J.~R., and Wikle, C.~K. (2020).
\newblock Ensemble {K}alman methods for high-dimensional hierarchical dynamic
  space-time models.
\newblock {\em Journal of the American Statistical Association},
  115(530):866--885.

\bibitem[Kendall, 1948]{kendall1948generalized}
Kendall, D.~G. (1948).
\newblock On the generalized ``birth-and-death" process.
\newblock {\em The Annals of Mathematical Statistics}, 19(1):1--15.

\bibitem[King et~al., 2008]{king2008inapparent}
King, A.~A., Ionides, E.~L., Pascual, M., and Bouma, M.~J. (2008).
\newblock Inapparent infections and cholera dynamics.
\newblock {\em Nature}, 454(7206):877--880.

\bibitem[King et~al., 2016]{king2015statistical}
King, A.~A., Nguyen, D., and Ionides, E.~L. (2016).
\newblock Statistical inference for partially observed {M}arkov processes via
  the {R} package pomp.
\newblock {\em Journal of Statistical Software}, 69(12).

\bibitem[Klugman et~al., 2013]{klugman2013loss}
Klugman, S.~A., Panjer, H.~H., and Willmot, G.~E. (2013).
\newblock {\em Loss models: {F}urther topics}.
\newblock John Wiley \& Sons.

\bibitem[Krak et~al., 2017]{krak2017efficient}
Krak, T., De~Bock, J., and Siebes, A. (2017).
\newblock Efficient computation of updated lower expectations for imprecise
  continuous-time hidden {M}arkov chains.
\newblock In {\em Proceedings of the Tenth International Symposium on Imprecise
  Probability: Theories and Applications}, pages 193--204. Proceedings of
  Machine Learning Research.

\bibitem[Li and Sendova, 2020]{li2020surplus}
Li, Y. and Sendova, K.~P. (2020).
\newblock A surplus process involving a compound poisson counting process and
  applications.
\newblock {\em Communications in Statistics-Theory and Methods},
  49(13):3238--3256.

\bibitem[Maraghi et~al., 2009]{maraghi2009batch}
Maraghi, F.~A., Madan, K.~C., and Darby-Dowman, K. (2009).
\newblock Batch arrival queueing system with random breakdowns and {B}ernoulli
  schedule server vacations having general vacation time distribution.
\newblock {\em International Journal of Information and Management Sciences},
  20(1):55--70.

\bibitem[Ning and Ionides, 2022]{ning2022iterated}
Ning, N. and Ionides, E.~L. (2022).
\newblock Iterated block particle filter for high-dimensional parameter
  learning: {B}eating the curse of dimensionality.
\newblock {\em in revision, Journal of Machine Learning Research}.

\bibitem[Ning and Wu, 2021]{ning2021well}
Ning, N. and Wu, J. (2021).
\newblock Well-posedness and stability analysis of two classes of generalized
  stochastic volatility models.
\newblock {\em SIAM Journal on Financial Mathematics}, 12(1):79--109.

\bibitem[Parthasarathy, 2005]{parthasarathy2005probability}
Parthasarathy, K.~R. (2005).
\newblock {\em Probability measures on metric spaces}, volume 352.
\newblock American Mathematical Soc.

\bibitem[Ramsay and Hooker, 2017]{ramsay2017dynamic}
Ramsay, J. and Hooker, G. (2017).
\newblock {\em Dynamic data analysis}.
\newblock Springer.

\bibitem[Sendova and Minkova, 2018]{sendova2018poisson}
Sendova, K.~P. and Minkova, L.~D. (2018).
\newblock Poisson-logarithmic risk process and applications.
\newblock {\em Comptes rendus de l’Acad{\'e}mie bulgare des Sciences}, 71(8).

\bibitem[Sendova and Minkova, 2020]{sendova2020introducing}
Sendova, K.~P. and Minkova, L.~D. (2020).
\newblock Introducing the non-homogeneous compound-birth process.
\newblock {\em Stochastics}, 92(5):814--832.

\bibitem[Zhang and Li, 2016]{zhang2016characterizations}
Zhang, H. and Li, B. (2016).
\newblock Characterizations of discrete compound {P}oisson distributions.
\newblock {\em Communications in Statistics-Theory and Methods},
  45(22):6789--6802.

\end{thebibliography}

\newpage
\begin{center}
	{\large\bf SUPPLEMENTARY MATERIAL}
\end{center}

\setcounter{section}{0}
\setcounter{page}{1}
\renewcommand{\thepage}{S\arabic{page}}
\renewcommand{\thesection}{S\arabic{section}}

\section{Illustration with integrated dispersion}
\label{Section:integrated}
The integrated dispersion index is another arrow-based measure of dispersion, whose ratio-formed definition in the current context is the following:
\begin{definition}
	\label{def:integrated_dispersion}
	For arrow $(v,v')\in A$, define the integrated variance 
	$$[\sigma_{vv'}^{\X}(t,\x_0)]^2:=\Var[N^{\X}_{vv'}(t)-N^{\X}_{vv'}(0)\mid \X(0)=\x_0],$$
	and the integrated mean 
	$$\mu_{vv'}^{\X}(t,\x_0):=\mathbb{E}[N^{\X}_{vv'}(t)-N^{\X}_{vv'}(0)\mid \X(0)=\x_0].$$
	Define the integrated dispersion index as the following ratio if it exists:
	$$D_{vv'}^{\X}(t,\x_0):=[\sigma_{vv'}^{\X}(t,\x_0)]^2\big /\mu_{vv'}^{\X}(t,\x_0).$$
	We say that with respect to arrow $(v,v')$, $\X(t)$ has integrated equi-dispersion if $D_{vv'}^{\X}(t,\x_0)=1$, has integrated over-dispersion if $D_{vv'}^{\X}(t,\x_0)>1$, and has integrated under-dispersion if $D_{vv'}^{\X}(t,\x_0)<1$.
\end{definition}

In the following three examples, we will show that no matter whether a model has integrated equi-dispersion (Example \ref{example_0}), under-dispersion (Example \ref{example_1}) or over-dispersion (Example \ref{example_2}), it can always have IED, with respect to a specific arrow. Throughout this section, we denote the GDM having IED as $\Z$.
\begin{example}
	\label{example_0}
	Consider the dynamic over arrow $(\overline{v},v)$ having the transition rate
	\begin{equation*}
		Q(t,(z_{\overline{v}},z_v,\z_{V\backslash \{\overline{v},v\}}),(z_{\overline{v}}-1,z_v+1, \z_{V\backslash \{\overline{v},v\}}))=\overline{r}_{\overline{v}v}(t)z_{\overline{v}}\mathbbm{1}_{\{z_{\overline{v}}\geq 1\}},
	\end{equation*}
	where $Q$ is defined in \eqref{eqn:transition_rate_Q}, $\Z(t)=\z$, and $\overline{r}_{\overline{v}v}$ is a positive function uniformly continuous in $t$.
	Then flow through arrow $(\overline{v},v)$ can be modeled by a Poisson process with intensity function $r_{\overline{v}v}$.
	The integrated mean and the integrated variance are given by 
	$$\mu_{\overline{v}v}^{\Z}(t,\z_0)=[\sigma_{\overline{v}v}^{\Z}(t,\z_0)]^2=\int_{0}^{t}r_{\overline{v}v}(s)ds,$$ 
	where $\Z(0)=\z_0$,
	and then the integrated dispersion index $D_{\overline{v},v}^{\Z}(t,\z_0)=1$; the infinitesimal mean and the infinitesimal variance are given by 
	$$\mu_{\overline{v}v}^{d\Z}(t,\z)=[\sigma_{\overline{v}v}^{d\Z}(t,\z)]^2=r_{\overline{v}v}(t),$$
	and then the infinitesimal dispersion index $D_{\overline{v},v}^{d\Z}(t,\z)=1$. Therefore, with respect to arrow $(\overline{v},v)$,  $\Z(t)$ has integrated equi-dispersion and has IED at $\Z(t)=\z$.
\end{example}

\begin{example}
	\label{example_1}
	When some components of $\Z(t)$ have non-negativity constraints, for example the number of individuals in each of the compartments must be non-negative at all times, modeling with unbounded processes is inappropriate. 
	We still consider the dynamic over arrow $(\overline{v},v)$ as Example \ref{example_0}, but under the boundedness constraint  
	we model 
	the flow over arrow $(\overline{v},v)$ by the time-inhomogeneous cumulative death process\footnote{The time-inhomogeneous cumulative death process is a MCP associated with a linear
		death process having individual death rate $\delta(t)$ and initial population size $d_0\in \mathbb{N}$, with transition rate $q(t,m,1)=\delta(t)(d_0-m)\mathbbm{1}_{\{m<d_0\}}$ and $q(t,m,k)=0$ for $k>1$.}. By \cite{kendall1948generalized} and Definition \ref{def:integrated_dispersion}, the integrated mean is given by
	$$\mu_{\overline{v}v}^{\Z}(t,\z_0)=z_{\overline{v}}^0 \left(1-e^{-\int_{0}^{t}\overline{r}_{\overline{v}v}(s)ds}\right),$$
	where $\Z(0)=\z_0$ and $Z_{\overline{v}}(0)=z_{\overline{v}}^0$, 
	the integrated variance is given by
	$$[\sigma_{\overline{v}v}^{\Z}(t,\z_0)]^2
	=z_{\overline{v}}^0 e^{-\int_{0}^{t}\overline{r}_{\overline{v}v}(s)ds}\left[1-e^{-\int_{0}^{t}\overline{r}_{\overline{v}v}(s)ds}\right]
	,$$
	and then the integrated dispersion index is given by
	$D_{\overline{v}v}^{\Z}(t,\z_0)=e^{-\int_{0}^{t}\overline{r}_{\overline{v}v}(s)ds}.$
	Given $\Z(t)=\z$, by L'H{\^o}pital's rule and Definition \ref{def:infinitesimal_dispersion}, the infinitesimal mean and the infinitesimal variance are given by
	$$\mu_{\overline{v}v}^{d\Z}(t,\z)=[\sigma_{\overline{v}v}^{d\Z}(t,\z)]^2=z_{\overline{v}}\overline{r}_{\overline{v}v}(t),$$
	and then we have the infinitesimal dispersion index 
	$D_{\overline{v}v}^{d\Z}(t,\z)=1.$
	Therefore, with respect to arrow $(\overline{v},v)$, $\Z(t)$ has integrated under-dispersion but has IED at $\Z(t)=\z$.
\end{example}

\begin{example}
	\label{example_2}
	Consider the dynamic over arrow $(\overline{v},v)$ having the transition rate
	\begin{equation*}
		Q(t,(z_{\overline{v}},z_v,\z_{V\backslash \{\overline{v},v\}}),(z_{\overline{v}}-1,z_v+1, \z_{V\backslash \{\overline{v},v\}}))=r_{\overline{v}v}(t)z_v\mathbbm{1}_{\{z_v\geq 1\}},
	\end{equation*}
	where $Q$ is defined in \eqref{eqn:transition_rate_Q}, $\Z(t)=\z$, and $r_{\overline{v}v}$ is a positive function uniformly continuous in $t$.
	The flow over arrow $(\overline{v},v)\in A$ can be modeled by the time-inhomogeneous birth process. 
	By \cite{kendall1948generalized} and Definition \ref{def:integrated_dispersion}, the integrated  mean is given by
	$$\mu_{\overline{v}v}^{\Z}(t,\z_0)=z_v^0 \left(e^{\int_{0}^{t}r_{\overline{v}v}(s)ds}-1\right),$$
	where $\Z(0)=\z_0$ and $Z_{v}(0)=z_{v}^0$, 
	the integrated variance is given by
	$$[\sigma_{\overline{v}v}^{\Z}(t,\z_0)]^2=z_v^0 e^{\int_{0}^{t}r_{\overline{v}v}(s)ds}\left[e^{\int_{0}^{t}r_{\overline{v}v}(s)ds}-1\right]
	,$$
	and then the integrated dispersion index is  given by
	$D_{\overline{v}v}^{\Z}(t,\z_0)=e^{\int_{0}^{t}r_{\overline{v}v}(s)ds}.$
	Given $\Z(t)=\z$, by L'H{\^o}pital's rule and Definition \ref{def:infinitesimal_dispersion},
	$$\mu_{\overline{v}v}^{d\Z}(t,\z)=[\sigma_{\overline{v}v}^{d\Z}(t,\z)]^2=z_v r_{\overline{v}v}(t),$$
	and then the infinitesimal dispersion index 
	is given by $D_{\overline{v}v}^{d\Z}(t,\z)=1$.
	Therefore, with respect to arrow $(\overline{v},v)$, $\Z(t)$ has integrated over-dispersion but has IED at $\Z(t)=\z$.
\end{example}

\section{Proof of Propositions}
\label{Section:proofs}

In this section, we give the proof of Proposition \ref{prop:iod_multinomial}
in Section \ref{Section:proof_multinomial}, and the proof of Proposition \ref{prop:iod_negative_multinomial}
in \ref{Section:proof_negative_multinomial}. We heavily use  properties of the multinomial distribution, the negative multinomial distritbution, and the Dirichlet distribution. We conduct asymptotic analysis in a sufficiently small time interval $[t,t+h]$. We recall that  increments over the arrow $(v,v')$ is defined as 
$$\Delta^{\X}_{vv'}(t,h)=N^{\X}_{vv'}(t+h)-N^{\X}_{vv'}(t).$$
We recall that for random variables $Y_1$, $Y_2$, and $Y_3$, 
the law of total variance states that 
$$\operatorname {Var} (Y_1)=\mathbb{E} [\operatorname {Var} (Y_1\mid Y_2)]+\operatorname {Var} (\mathbb{E} [Y_1\mid Y_2]),$$
and the law of total covariance
states that 
$$\operatorname {cov} (Y_1,Y_2)=\mathbb{E} (\operatorname {cov} (Y_1,Y_2\mid Y_3))+\operatorname {cov} (\mathbb{E} (Y_1\mid Y_3),\mathbb{E} (Y_2\mid Y_3)).$$

\subsection{Proof of Proposition \ref{prop:iod_multinomial}}
\label{Section:proof_multinomial}

\noindent $(1)$. By \eqref{eqn:multinomial_pdf_modified}, for $i\in \{1,\ldots,m\}$
\begin{align*}
	\mathbb{E}(\Delta^{\X}_{vv_i'}(t,h)\mid \X(t)=\x)=x_v\mathbb{E}(\Pi_{vv_i'}(t,h,\x))+o(h)=\frac{x_v}{c}\alpha_{vv_i'}(t,h,\x)+o(h),
\end{align*}
and by the law of total variance
\begin{align*}
	&\hspace*{-0.1cm}\Var(\Delta^{\X}_{vv_i'}(t,h)\mid \X(t)=\x)\\
	&=x_v^2\Var(\Pi_{vv_i'}(t,h,\x))+x_v\mathbb{E}(\Pi_{vv_i'}(t,h,\x)(1-\Pi_{vv_i'}(t,h,\x)))+o(h)\\
	&=(x_v^2-x_v)\frac{\alpha_{vv_i'}(t,h,\x)(c-\alpha_{vv_i'}(t,h,\x))}{c^2(c+1)}+\frac{x_v}{c}\alpha_{vv_i'}(t,h,\x)\\
	&\quad\quad-\frac{x_v}{c^2}\left(\alpha_{vv_i'}(t,h,\x)\right)^2+o(h).
\end{align*}
Plugging 
$$\alpha_{vv_i'}(t,h,\x)=c\left(1-e^{-\sum_{j=1}^{m}\int_{t}^{t+h}r_{vv_j'}(s,\x)ds}\right)\frac{r_{vv_i'}(t,\x)}{\sum_{j=1}^{m}r_{vv_j'}(t,\x)}$$
into the above two equations, we have
\begin{align*}
	\mathbb{E}(\Delta^{\X}_{vv_i'}(t,h)\mid \X(t)=\x)=x_v\left(1-e^{-\sum_{j=1}^{m}\int_{t}^{t+h}r_{vv_j'}(s,\x)ds}\right)\frac{r_{vv_i'}(t,\x)}{\sum_{j=1}^{m}r_{vv_j'}(t,\x)}+o(h)
\end{align*}
and
\begin{align*}
	&\hspace*{-0.1cm}\Var(\Delta^{\X}_{vv_i'}(t,h)\mid \X(t)=\x)\\
	&=(x_v^2-x_v)(c+1)^{-1}\left(1-e^{-\sum_{j=1}^{m}\int_{t}^{t+h}r_{vv_j'}(s,\x)ds}\right)\frac{r_{vv_i'}(t,\x)}{\sum_{j=1}^{m}r_{vv_j'}(t,\x)}\\
	&\quad\quad\quad\quad\quad\times \left(1-\left(1-e^{-\sum_{j=1}^{m}\int_{t}^{t+h}r_{vv_j'}(s,\x)ds}\right)\frac{r_{vv_i'}(t,\x)}{\sum_{j=1}^{m}r_{vv_j'}(t,\x)}\right)\\
	&\quad\quad+x_v\left(1-e^{-\sum_{j=1}^{m}\int_{t}^{t+h}r_{vv_j'}(s,\x)ds}\right)\frac{r_{vv_i'}(t,\x)}{\sum_{j=1}^{m}r_{vv_j'}(t,\x)}\\
	&\quad\quad-x_v\left(\left(1-e^{-\sum_{j=1}^{m}\int_{t}^{t+h}r_{vv_j'}(s,\x)ds}\right)\frac{r_{vv_i'}(t,\x)}{\sum_{j=1}^{m}r_{vv_j'}(t,\x)}\right)^2+o(h).
\end{align*}
Note that when $h$ is sufficiently small, we have $\mathbb{E}(\Delta^{\X}_{vv_i'}(t,h)\mid \X(t)=\x)>0$ and
since
\begin{align*}
	&\hspace*{-1cm}\left(1-e^{-\sum_{j=1}^{m}\int_{t}^{t+h}r_{vv_j'}(s,\x)ds}\right)\frac{r_{vv_i'}(t,\x)}{\sum_{j=1}^{m}r_{vv_j'}(t,\x)}\\
	&>\left(\left(1-e^{-\sum_{j=1}^{m}\int_{t}^{t+h}r_{vv_j'}(s,\x)ds}\right)\frac{r_{vv_i'}(t,\x)}{\sum_{j=1}^{m}r_{vv_j'}(t,\x)}\right)^2,
\end{align*}
we also have 
$\Var(\Delta^{\X}_{vv_i'}(t,h)\mid \X(t)=\x)>0$.
Applying L'H{\^o}pital's rule,
\begin{align*}
	\mu_{vv_i'}^{d\X}(t,\x)=&x_v r_{vv_i'}(t,\x),\\
	[\sigma_{vv_i'}^{d\X}(t,\x)]^2=&(1+(x_v-1)(c+1)^{-1})x_v r_{vv_i'}(t,\x).
\end{align*}
Then we can see that for any $i\in \{1,\ldots,m\}$, when $x_v>1$, the infinitesimal dispersion index $D_{vv_i'}^{d\X}(t,\x)>1$, i.e., $\X(t)$ has IOD at $\X(t)=\x$ with respect to arrow $(v,v_i')\in A$; when $x_v=1$, the infinitesimal dispersion index $D_{vv_i'}^{d\X}(t,\x)=1$, i.e., $\X(t)$ has IED at $\X(t)=\x$ with respect to arrow $(v,v_i')\in A$.  By Definition \ref{def:system_infinitesimal_dispersion}, when $x_v>1$, $\X(t)$ has SIOD at $\X(t)=\x$ over connected outgoing arrows $\{(v,v_i')\}_{i\in \{1,\ldots,m\}}$; when $x_v=1$, $\X(t)$ has SIED at $\X(t)=\x$ over connected outgoing arrows $\{(v,v_i')\}_{i\in \{1,\ldots,m\}}$.

By the  law of total covariance, for $i,j\in \{1,\ldots,m\}$ and $i\neq j$, we have
\begin{align}
	\label{eqn:cov_negative_multinomial}
	&\Cov[\Delta^{\X}_{vv_i'}(t,h),\Delta^{\X}_{vv_j'}(t,h)\mid  \X(t)=\x]\nonumber\\
	&=\Cov\Big[\mathbb{E}[\Delta^{\X}_{vv_i'}(t,h)\mid  \X(t)=\x,\{\Pi_{vv_i'}(t,h,\x)\}_{i\in \{0,\ldots,m\}}],\nonumber\\
	&\hspace*{3.9cm}\mathbb{E}[\Delta^{\X}_{vv_j'}(t,h)\mid  \X(t)=\x,\{\Pi_{vv_i'}(t,h,\x)\}_{i\in \{0,\ldots,m\}}]\Big]\nonumber\\
	&\quad+\mathbb{E}\Big[\Cov[\Delta^{\X}_{vv_i'}(t,h),\Delta^{\X}_{vv_j'}(t,h)\mid  \X(t)=\x,\{\Pi_{vv_i'}(t,h,\x)\}_{i\in \{0,\ldots,m\}}]\Big]\nonumber\\
	&=x_{v}^2\Cov[\Pi_{vv_i'}(t,h,\x), \Pi_{vv_j'}(t,h,\x)]-\mathbb{E}[x_{v}\Pi_{vv_i'}(t,h,\x)\Pi_{vv_j'}(t,h,\x)]+o(h)\nonumber\\
	&=(x_{v}^2-x_{v})\Cov[\Pi_{vv_i'}(t,h,\x), \Pi_{vv_j'}(t,h,\x)]\nonumber\\
	&\quad-x_{v}\mathbb{E}[\Pi_{vv_i'}(t,h,\x)]\mathbb{E}[\Pi_{vv_j'}(t,h,\x)]+o(h)\nonumber\\
	&=-(x_{v}^2-x_{v})\frac{\alpha_{vv_i'}(t,h,\x)\alpha_{vv_j'}(t,h,\x)}{c^2(c+1)}-x_{v}\frac{\alpha_{vv_i'}(t,h,\x)\alpha_{vv_j'}(t,h,\x)}{c^2}+o(h)\nonumber\\
	&=-\frac{(x_{v}^2-x_{v})}{c+1}\left(1-e^{-\sum_{j=1}^{m}\int_{t}^{t+h}r_{vv_j'}(s,\x)ds}\right)^2\frac{r_{vv_i'}(t,\x)r_{vv_j'}(t,\x)}{(\sum_{j=1}^{m}r_{vv_j'}(t,\x))^2}\nonumber\\
	&\quad-x_{v}\left(1-e^{-\sum_{j=1}^{m}\int_{t}^{t+h}r_{vv_j'}(s,\x)ds}\right)^2\frac{r_{vv_i'}(t,\x)r_{vv_j'}(t,\x)}{(\sum_{j=1}^{m}r_{vv_j'}(t,\x))^2}+o(h).
\end{align}
By Definition \ref{def:infinitesimal_covariance}, for $i,j\in \{1,\ldots,m\}$ and $i\neq j$,
\begin{align*}
	\sigma_{vv_i',vv_j'}^{d\X}(t,\x)=0.
\end{align*}

\noindent $(2)$. 
By \eqref{eqn:multinomial_pdf_modified}, with $\mathbf{y}=(y_0,y_1,\ldots,y_m)$, we have
\begin{align}
	\label{eqn:noise_pdf_binomial}
	&\mathbb{P}(\{\Delta^{\X}_{vv_i'}(t,h)=k_{i}\}_{i\in \{0,\ldots,m\}} \mid \X(t)=\x)\nonumber\\
	&=\int_0^1\frac{\Gamma(x_v+1)}{\prod_{i=0}^{m}\Gamma(k_{i}+1)}\prod_{i=0}^{m}\left(y_i\right)^{k_{i}}\left\{\frac{\Gamma(\sum_{i=0}^{m}\alpha_{vv_i'}(t,h,\x) )}{\prod_{i=0}^{m}\Gamma(\alpha_{vv_i'}(t,h,\x))}\prod_{i=0}^{m}y_i^{\alpha_{vv_i'}(t,h,\x)-1}\right\} d\mathbf{y}\nonumber\\
	&\quad\quad+o(h)\nonumber\\
	&=\frac{\Gamma(x_v+1)}{\prod_{i=0}^{m}\Gamma(k_{i}+1)}\frac{\Gamma(\sum_{i=0}^{m}\alpha_{vv_i'}(t,h,\x) )}{\prod_{i=0}^{m}\Gamma(\alpha_{vv_i'}(t,h,\x))}\int\prod_{i=0}^{m}\left(y_i\right)^{k_{i\in |A|}+\alpha_{vv_i'}(t,h,\x)-1} d\mathbf{y}+o(h)\nonumber\\
	&=\frac{\Gamma(x_v+1)}{\prod_{i=0}^{m}\Gamma(k_{i}+1)}\frac{\Gamma(\sum_{i=0}^{m}\alpha_{vv_i'}(t,h,\x) )}{\prod_{i=0}^{m}\Gamma(\alpha_{vv_i'}(t,h,\x))}\frac{\prod_{i=0}^{m}\Gamma(k_{i}+\alpha_{vv_i'}(t,h,\x))}{\Gamma(\sum_{i=0}^{m}k_{i}+\sum_{i=0}^{m}\alpha_{vv_i'}(t,h,\x) )}+o(h)\nonumber\\
	&=\frac{\Gamma(x_v+1)}{\prod_{i=0}^{m}\Gamma(k_{i}+1)}\frac{\Gamma(c )}{\prod_{i=0}^{m}\Gamma(\alpha_{vv_i'}(t,h,\x))}\frac{\prod_{i=0}^{m}\Gamma(k_{i}+\alpha_{vv_i'}(t,h,\x))}{\Gamma(x_v+c )}+o(h).
\end{align}
Recalling that for $i\in \{1,\ldots,m\}$
\begin{align*}
	\alpha_{vv_i'}(t,h,\x)
	=&c\left(1-e^{-\sum_{j=1}^{m}\int_{t}^{t+h}r_{vv_j'}(s,\x)ds}\right)\frac{r_{vv_i'}(t,\x)}{\sum_{j=1}^{m}r_{vv_j'}(t,\x)},
\end{align*}
by Taylor series we have
\begin{align}
	\label{eqn:property0_multinomial}
	\alpha_{vv_i'}(t,h,\x)=cr_{vv_i'}(t,\x) h+o(h).
\end{align}
Hence, for $k_{i}\geq 1$ and $i\in \{1,\ldots,m\}$, 
\begin{align}
	\label{eqn:property1_multinomial}
	\Gamma(k_i+\alpha_{vv_i'}(t,h,\x))
	=&(k_i+\alpha_{vv_i'}(t,h,\x)-1)\cdots (2+\alpha_{vv_i'}(t,h,\x)-1)\nonumber\\
	&\times\alpha_{vv_i'}(t,h,\x)\cdot\Gamma(\alpha_{vv_i'}(t,h,\x))\nonumber\\
	=&(k_i-1)\cdots (2-1)\cdot\alpha_{vv_i'}(t,h,\x)\cdot\Gamma(\alpha_{vv_i'}(t,h,\x))+o(\alpha_{vv_i'}(t,h,\x))\nonumber\\
	=&\Gamma(k_i)\alpha_{vv_i'}(t,h,\x)\Gamma(\alpha_{vv_i'}(t,h,\x))+o(\alpha_{vv_i'}(t,h,\x)).
\end{align}
Furthermore, we have that 
\begin{align}
	\label{eqn:property2_multinomial}
	&\Gamma(k_{0}+\alpha_{vv_0'}(t,h,\x))\nonumber\\
	&=\Gamma\left(x_v-\sum_{i=1}^{m}k_{i\in |A|}+c-\sum_{i=1}^{m}\alpha_{vv_i'}(t,h,\x)\right)\nonumber\\
	&=\left(x_v-\sum_{i=1}^{m}k_{i\in |A|}-1+c-\sum_{i=1}^{m}\alpha_{vv_i'}(t,h,\x)\right)\cdots \left(c-\sum_{i=1}^{m}\alpha_{vv_i'}(t,h,\x)\right)\nonumber\\
	&\quad\quad\times \Gamma(\alpha_{vv_0'}(t,h,\x))\nonumber\\
	&=\left[\left(x_v-\sum_{i=1}^{m}k_{i\in |A|}-1+c\right)\cdots c+\mathcal{O}\left(\sum_{i=1}^{m}\alpha_{vv_i'}(t,h,\x)\right)\right]\Gamma(\alpha_{vv_0'}(t,h,\x))\nonumber\\
	&=\left[\frac{\Gamma(x_v-\sum_{i=1}^{m}k_{i\in |A|}+c)}{\Gamma(c)}+\mathcal{O}\left(\sum_{i=1}^{m}\alpha_{vv_i'}(t,h,\x)\right)\right]\Gamma(\alpha_{vv_0'}(t,h,\x)).
\end{align}
Plugging \eqref{eqn:property0_multinomial}, \eqref{eqn:property1_multinomial}, and \eqref{eqn:property2_multinomial} into \eqref{eqn:noise_pdf_binomial}, we can see that
\begin{align*}
	\mathbb{P}(\{\Delta^{\X}_{vv_i'}(t,h)=k_{i}\}_{i\in \{0,1,\ldots,m\}},\; |\mathcal{S}|\geq 2  \mid \X(t)=\x)=o(h)
\end{align*}
and
\begin{align*}
	\mathbb{P}(\{\Delta^{\X}_{vv_i'}(t,h)=k_{i}\}_{i\in \{0,1,\ldots,m\}},\; |\mathcal{S}|=1  \mid \X(t)=\x)=\sum_{i=1}^{m}q_{{v}{v_i'}}(t,\x,k_i)h+o(h),
\end{align*}
where $\mathcal{S}$ is the set defined in \eqref{eqn:setA_multinomial}, $|\mathcal{S}|$ is the cardinality of $\mathcal{S}$, and for $i\in \{1,\ldots,m\}$
\begin{align*}
	q_{{v}{v_i'}}(t,\x,k_i)
	=c{x_v \choose k_i}\frac{\Gamma(k_i)\Gamma(x_v-k_i+c)}{\Gamma(x_v+c)}r_{vv_i'}(t,\x).
\end{align*}

\subsection{Proof of Proposition \ref{prop:iod_negative_multinomial}}
\label{Section:proof_negative_multinomial}
\noindent $(1)$. By \eqref{eqn:negative_multinomial_pdf_modified}, for $i\in \{1,\ldots,\overline{m}\}$, with $\mathbf{y}=(y_0,y_1,\ldots,y_{\overline{m}})$, we have
\begin{align}
	\label{eqn:negative_multinomial_mean}
	&\mathbb{E}(\Delta^{\X}_{u_iu'}(t,h)\mid \X(t)=\x)\nonumber\\
	&=x_{u'}\mathbb{E}\left((\Pi_{u_0 u'}(t,h,\x))^{-1}\Pi_{u_iu'}(t,h,\x)\right)+o(h)\nonumber\\
	&=x_{u'}\int_{0}^{1}y_0^{-1}y_i\frac{\Gamma(\sum_{i=0}^{\overline{m}}\alpha_{u_iu'}(t,h,\x))}{\prod_{i=0}^{\overline{m}}\Gamma(\alpha_{u_iu'}(t,h,\x))} \prod_{i=0}^{\overline{m}}y_i^{\alpha_{u_iu'}(t,h,\x)-1}d\y+o(h)\nonumber\\
	&=\int_{0}^{1}y_0^{\alpha_{u_0 u'}(t,h,\x)-2}y_i^{\alpha_{u_iu'}(t,h,\x)}\prod_{j=1}^{{i-1}}y_j^{\alpha_{u_ju'}(t,h,\x)-1}
	\prod_{j=i+1}^{\overline{m}}y_j^{\alpha_{u_ju'}(t,h,\x)-1}d\y \nonumber\\
	&\quad\times x_{u'}\frac{\Gamma(\sum_{i=0}^{\overline{m}}\alpha_{u_iu'}(t,h,\x))}{\prod_{i=0}^{\overline{m}}\Gamma(\alpha_{u_iu'}(t,h,\x))}+o(h)\nonumber\\
	&=\frac{\Gamma(\sum_{i=0}^{\overline{m}}\alpha_{u_iu'}(t,h,\x))}{\prod_{i=0}^{\overline{m}}\Gamma(\alpha_{u_iu'}(t,h,\x))}
	\frac{\prod_{j=1}^{i-1}\Gamma(\alpha_{u_ju'}(t,h,\x))\prod_{j=i+1}^{\overline{m}}\Gamma(\alpha_{u_ju'}(t,h,\x))}{\Gamma(\sum_{i=0}^{\overline{m}}\alpha_{u_iu'}(t,h,\x))}
	\nonumber\\
	&\quad\times x_{u'}\Gamma(\alpha_{u_0 u'}(t,h,\x)-1)\Gamma(\alpha_{u_iu'}(t,h,\x)+1)+o(h)\nonumber\\
	&=x_{u'}\frac{\Gamma(\alpha_{u_0 u'}(t,h,\x)-1)\Gamma(\alpha_{u_iu'}(t,h,\x)+1)}{\Gamma(\alpha_{u_0 u'}(t,h,\x))\Gamma(\alpha_{u_iu'}(t,h,\x))}+o(h)\nonumber\\
	&=x_{u'}\frac{\alpha_{u_iu'}(t,h,\x)}{\alpha_{u_0 u'}(t,h,\x)-1}+o(h).
\end{align}
By \eqref{eqn:negative_multinomial_pdf_modified} and the law of total variance, we have
\begin{align}
	\label{eqn:negative_multinomial_variance}
	&\Var(\Delta^{\X}_{u_iu'}(t,h)\mid \X(t)=\x)\nonumber\\
	=&x_{u'}^2\Var((\Pi_{u_0 u'}(t,h,\x))^{-1}\Pi_{u_iu'}(t,h,\x))+x_{u'}\mathbb{E}[(\Pi_{u_0 u'}(t,h,\x))^{-2}(\Pi_{u_iu'}(t,h,\x))^2]\nonumber\\
	&+x_{u'}\mathbb{E}[(\Pi_{u_0 u'}(t,h,\x))^{-1}\Pi_{u_iu'}(t,h,\x)]+o(h)\nonumber\\
	=&(x_{u'}^2+x_{u'})\mathbb{E}[(\Pi_{u_0 u'}(t,h,\x))^{-2}(\Pi_{u_iu'}(t,h,\x))^2]-x_{u'}^2\mathbb{E}[(\Pi_{u_0 u'}(t,h,\x))(\Pi_{u_iu'}(t,h,\x))]^2\nonumber\\
	&+x_{u'}\mathbb{E}[(\Pi_{u_0 u'}(t,h,\x))^{-1}\Pi_{u_iu'}(t,h,\x)]+o(h)\nonumber\\
	=&(x_{u'}^2+x_{u'})\frac{\Gamma(\alpha_{u_0 u'}(t,h,\x)-2)\Gamma(\alpha_{u_iu'}(t,h,\x)+2)}{\Gamma(\alpha_{u_0 u'}(t,h,\x))\Gamma(\alpha_{u_iu'}(t,h,\x))}\nonumber\\
	&-x_{u'}^2\left(\frac{\Gamma(\alpha_{u_0 u'}(t,h,\x)-1)\Gamma(\alpha_{u_iu'}(t,h,\x)+1)}{\Gamma(\alpha_{u_0 u'}(t,h,\x))\Gamma(\alpha_{u_iu'}(t,h,\x))}\right)^2\nonumber\\
	&+x_{u'}\left(\frac{\Gamma(\alpha_{u_0 u'}(t,h,\x)-1)\Gamma(\alpha_{u_iu'}(t,h,\x)+1)}{\Gamma(\alpha_{u_0 u'}(t,h,\x))\Gamma(\alpha_{u_iu'}(t,h,\x))}\right)+o(h)\nonumber\\
	=&(x_{u'}^2+x_{u'})\frac{(\alpha_{u_iu'}(t,h,\x)+1)\alpha_{u_iu'}(t,h,\x)}{(\alpha_{u_0 u'}(t,h,\x)-1)(\alpha_{u_0 u'}(t,h,\x)-2)}\nonumber\\
	&-x_{u'}^2\left(\frac{\alpha_{u_iu'}(t,h,\x)}{\alpha_{u_0 u'}(t,h,\x)-1}\right)^2+x_{u'}\left(\frac{\alpha_{u_iu'}(t,h,\x)}{\alpha_{u_0 u'}(t,h,\x)-1}\right)+o(h).
\end{align}

Plugging 
\begin{align}
	\label{eqn:alpha_negativemultinomial}
	\alpha_{u_iu'}(t,h,\x)=c\left(1-e^{-\sum_{j=1}^{\overline{m}}\int_{t}^{t+h}r_{u_j u'}(s,\x)ds}\right)\frac{r_{u_i u'}(t,\x)}{\sum_{j=1}^{\overline{m}}r_{u_j u'}(t,\x)}
\end{align}
and 
\begin{align}
	\label{eqn:alpha0_negativemultinomial}
	\alpha_{u_0 u'}(t,h,\x)=c-\sum_{i=1}^{m}\alpha_{u_iu'}(t,h,\x)=ce^{-\sum_{j=1}^{\overline{m}}\int_{t}^{t+h}r_{u_j u'}(s,\x)ds}
\end{align}
into equations \eqref{eqn:negative_multinomial_mean} and \eqref{eqn:negative_multinomial_variance}, we have
\begin{align}
	\label{eqn:mean_increment_multibirth}
	\mathbb{E}(\Delta^{\X}_{u_iu'}(t,h)\mid \X(t)=\x)=x_{u'}\frac{c\left(1-e^{-\sum_{j=1}^{\overline{m}}\int_{t}^{t+h}r_{u_j u'}(s,\x)ds}\right)\frac{r_{u_i u'}(t,\x)}{\sum_{j=1}^{\overline{m}}r_{u_j u'}(t,\x)}}{ce^{-\sum_{j=1}^{\overline{m}}\int_{t}^{t+h}r_{u_j u'}(s,\x)ds}-1}+o(h)
\end{align}
and
\begin{align}
	\label{eqn:variance_increment_multibirth}
	&\hspace*{-0.2cm}\Var(\Delta^{\X}_{u_iu'}(t,h)\mid \X(t)=\x)\nonumber\\
	=&(x_{u'}^2+x_{u'})\frac{c\left(1-e^{-\sum_{j=1}^{\overline{m}}\int_{t}^{t+h}r_{u_j u'}(s,\x)ds}\right)\frac{r_{u_i u'}(t,\x)}{\sum_{j=1}^{\overline{m}}r_{u_j u'}(t,\x)}+1}{\left(ce^{-\sum_{j=1}^{\overline{m}}\int_{t}^{t+h}r_{u_j u'}(s,\x)ds}-1\right)\left(ce^{-\sum_{j=1}^{\overline{m}}\int_{t}^{t+h}r_{u_j u'}(s,\x)ds}-2\right)}\nonumber\\
	&\quad\quad\times c\left(1-e^{-\sum_{j=1}^{\overline{m}}\int_{t}^{t+h}r_{u_j u'}(s,\x)ds}\right)\frac{r_{u_i u'}(t,\x)}{\sum_{j=1}^{\overline{m}}r_{u_j u'}(t,\x)}\nonumber\\
	&-x_{u'}^2\left(\frac{c\left(1-e^{-\sum_{j=1}^{\overline{m}}\int_{t}^{t+h}r_{u_j u'}(s,\x)ds}\right)\frac{r_{u_i u'}(t,\x)}{\sum_{j=1}^{\overline{m}}r_{u_j u'}(t,\x)}}{ce^{-\sum_{j=1}^{\overline{m}}\int_{t}^{t+h}r_{u_j u'}(s,\x)ds}-1}\right)^2\nonumber\\
	&+x_{u'}\left(\frac{c\left(1-e^{-\sum_{j=1}^{\overline{m}}\int_{t}^{t+h}r_{u_j u'}(s,\x)ds}\right)\frac{r_{u_i u'}(t,\x)}{\sum_{j=1}^{\overline{m}}r_{u_j u'}(t,\x)}}{ce^{-\sum_{j=1}^{\overline{m}}\int_{t}^{t+h}r_{u_j u'}(s,\x)ds}-1}\right)+o(h).
\end{align}
Note that when $h$ is sufficiently small, $c>2e^{\sum_{i=1}^{\overline{m}}\int_{t}^{t+h}r_{u_iu'}(s,\x)ds}$ suffices that $\mathbb{E}(\Delta^{\X}_{u_iu'}(t,h)\mid \X(t)=\x)>0$,
and together with the fact that
\begin{align*}
	\frac{\left(1-e^{-\sum_{j=1}^{\overline{m}}\int_{t}^{t+h}r_{u_j u'}(s,\x)ds}\right)\frac{r_{u_i u'}(t,\x)}{\sum_{j=1}^{\overline{m}}r_{u_j u'}(t,\x)}+1}{ce^{-\sum_{j=1}^{\overline{m}}\int_{t}^{t+h}r_{u_j u'}(s,\x)ds}-2}>\frac{c\left(1-e^{-\sum_{j=1}^{\overline{m}}\int_{t}^{t+h}r_{u_j u'}(s,\x)ds}\right)\frac{r_{u_i u'}(t,\x)}{\sum_{j=1}^{\overline{m}}r_{u_j u'}(t,\x)}}{ce^{-\sum_{j=1}^{\overline{m}}\int_{t}^{t+h}r_{u_j u'}(s,\x)ds}-1}
\end{align*}
we also have 
$\Var(\Delta^{\X}_{u_iu'}(t,h)\mid \X(t)=\x)>0$.

By L'H{\^o}pital's rule, equations \eqref{eqn:mean_increment_multibirth} and \eqref{eqn:variance_increment_multibirth}, and Definition \ref{def:infinitesimal_dispersion}, we have
\begin{align*}
	\mu_{u_i u'}^{d\X}(t,\x)=&x_{u'} r_{u_iu'}(t,\x)\frac{c}{c-1},\\
	[\sigma_{u_i u'}^{d\X}(t,\x)]^2=&x_{u'}^2 r_{u_iu'}(t,\x) \frac{c}{(c-1)(c-2)}+x_{u'} r_{u_iu'}(t,\x) \frac{c}{c-2}.
\end{align*}
When $c>2e^{\sum_{i=1}^{\overline{m}}\int_{t}^{t+h}r_{u_iu'}(s,\x)ds}$, since 
$$x_{u'} r_{u_iu'}(t,\x) \frac{c}{c-2}>x_{u'} r_{u_iu'}(t,\x)\frac{c}{c-1}>0$$
and 
$$x_{u'}^2 r_{u_iu'}(t,\x) \frac{c}{(c-1)(c-2)}>0,$$
we have $D_{u_iu'}^{d\X}(t,\x)>1$, i.e., $\X(t)$ has IOD at $\X(t)=\x$ with respect to each arrow in $\{(u_i, u')\}_{i\in \{1,\ldots,\overline{m}\}}$. By Definition \ref{def:system_infinitesimal_dispersion}, $\X(t)$ has SIOD at $\X(t)=\x$ over connected incoming arrows $\{(u_i, u')\}_{i\in \{1,\ldots,\overline{m}\}}$.

By the  law of total covariance, for $i,j\in \{1,\ldots,\overline{m}\}$ and $i\neq j$, we have
\begin{align}
	\label{eqn:cov_negative_multinomial}
	&\Cov[\Delta^{\X}_{u_iu'}(t,h),\Delta^{\X}_{u_ju'}(t,h)\mid  \X(t)=\x]\nonumber\\
	&=\Cov[\mathbb{E}[\Delta^{\X}_{u_iu'}(t,h)\mid  \X(t)=\x,\{\Pi_{u_i u'}(t,h,\x)\}_{i\in \{0,1,\ldots,\overline{m}\}}],\nonumber\\
	&\hspace*{3.9cm}\mathbb{E}[\Delta^{\X}_{u_ju'}(t,h)\mid  \X(t)=\x,\{\Pi_{u_i u'}(t,h,\x)\}_{i\in \{0,1,\ldots,\overline{m}\}}]]\nonumber\\
	&\quad+\mathbb{E}[\Cov[\Delta^{\X}_{u_iu'}(t,h),\Delta^{\X}_{u_ju'}(t,h)\mid  \X(t)=\x,\{\Pi_{u_i u'}(t,h,\x)\}_{i\in \{0,1,\ldots,\overline{m}\}}]]+o(h)\nonumber\\
	&=x_{u'}^2\Cov[(\Pi_{u_0 u'}(t,h,\x))^{-1}\Pi_{u_iu'}(t,h,\x), (\Pi_{u_0 u'}(t,h,\x))^{-1}\Pi_{u_ju'}(t,h,\x)]\nonumber\\
	&\quad+\mathbb{E}[x_{u'}(\Pi_{u_0 u'}(t,h,\x))^{-2}\Pi_{u_iu'}(t,h,\x)\Pi_{u_ju'}(t,h,\x)]+o(h)\nonumber\\
	&=x_{u'}^2\mathbb{E}[(\Pi_{u_0 u'}(t,h,\x))^{-2}\Pi_{u_iu'}(t,h,\x)\Pi_{u_ju'}(t,h,\x)]\nonumber\\
	&\quad-x_{u'}^2\mathbb{E}[(\Pi_{u_0 u'}(t,h,\x))^{-1}\Pi_{u_iu'}(t,h,\x)]\mathbb{E}[(\Pi_{u_0 u'}(t,h,\x))^{-1}\Pi_{u_ju'}(t,h,\x)]\nonumber\\
	&\quad+x_{u'}\mathbb{E}[(\Pi_{u_0 u'}(t,h,\x))^{-2}\Pi_{u_iu'}(t,h,\x)\Pi_{u_ju'}(t,h,\x)]+o(h)\nonumber\\
	&=(x_{u'}^2+x_{u'})\mathbb{E}[(\Pi_{u_0 u'}(t,h,\x))^{-2}\Pi_{u_iu'}(t,h,\x)\Pi_{u_ju'}(t,h,\x)]\nonumber\\
	&\quad-x_{u'}^2\mathbb{E}[(\Pi_{u_0 u'}(t,h,\x))^{-1}\Pi_{u_iu'}(t,h,\x)]\mathbb{E}[(\Pi_{u_0 u'}(t,h,\x))^{-1}\Pi_{u_ju'}(t,h,\x)]+o(h).
\end{align}
Note that
\begin{align}
	\label{eqn:cov_negative_multinomial0}
	&\mathbb{E}[(\Pi_{u_0 u'}(t,h,\x))^{-2}\Pi_{u_iu'}(t,h,\x)\Pi_{u_ju'}(t,h,\x)]\nonumber\\
	&=\int_{0}^{1}y_0^{-2}y_iy_j\frac{\Gamma(\sum_{i=0}^{\overline{m}}\alpha_{u_iu'}(t,h,\x))}{\prod_{i=0}^{\overline{m}}\Gamma(\alpha_{u_iu'}(t,h,\x))} \prod_{i=0}^{\overline{m}}y_i^{\alpha_{u_iu'}(t,h,\x)-1}d\y+o(h)\nonumber\\
	&=\frac{\Gamma(\sum_{i=0}^{\overline{m}}\alpha_{u_iu'}(t,h,\x))}{\prod_{i=0}^{\overline{m}}\Gamma(\alpha_{u_iu'}(t,h,\x))} \int_{0}^{1}y_0^{\alpha_{u_0 u'}(t,h,\x)-3}y_i^{\alpha_{u_iu'}(t,h,\x)}y_j^{\alpha_{u_ju'}(t,h,\x)}\nonumber\\
	&\hspace*{5cm}\times\prod_{l\in \{1,\ldots,\overline{m}\}\backslash  \{i,j\}}^{}y_l^{\alpha_{u_lu'}(t,h,\x)-1}d\y+o(h)\nonumber\\
	&=\Gamma(\alpha_{u_0 u'}(t,h,\x)-2)\Gamma(\alpha_{u_iu'}(t,h,\x)+1)\Gamma(\alpha_{u_ju'}(t,h,\x)+1)\nonumber\\
	&\hspace*{0.5cm}\times\frac{\Gamma(\sum_{i=0}^{\overline{m}}\alpha_{u_iu'}(t,h,\x))}{\prod_{i=0}^{\overline{m}}\Gamma(\alpha_{u_iu'}(t,h,\x))} \frac{\prod_{l\in \{1,\ldots,\overline{m}\}\backslash  \{i,j\}}^{}\Gamma(\alpha_{u_lu'}(t,h,\x))}{\Gamma(\sum_{i=0}^{\overline{m}}\alpha_{u_iu'}(t,h,\x))}+o(h)\nonumber\\
	&=\frac{\Gamma(\alpha_{u_0 u'}(t,h,\x)-2)\Gamma(\alpha_{u_iu'}(t,h,\x)+1)\Gamma(\alpha_{u_ju'}(t,h,\x)+1)}{\Gamma(\alpha_{u_0 u'}(t,h,\x))\Gamma(\alpha_{u_iu'}(t,h,\x))\Gamma(\alpha_{u_ju'}(t,h,\x))}+o(h)\nonumber\\
	&=\frac{\alpha_{u_iu'}(t,h,\x)\alpha_{u_ju'}(t,h,\x)}{(\alpha_{u_0 u'}(t,h,\x)-1)(\alpha_{u_0 u'}(t,h,\x)-2)}+o(h).
\end{align}
Plugging equations \eqref{eqn:alpha_negativemultinomial} and \eqref{eqn:alpha0_negativemultinomial} into \eqref{eqn:cov_negative_multinomial0}, we have
\begin{align}
	\label{eqn:cov_negative_multinomial1}
	&\mathbb{E}[(\Pi_{u_0 u'}(t,h,\x))^{-2}\Pi_{u_iu'}(t,h,\x)\Pi_{u_ju'}(t,h,\x)]\nonumber\\
	&=\frac{c^2\left(1-e^{-\sum_{j=1}^{\overline{m}}\int_{t}^{t+h}r_{u_j u'}(s,\x)ds}\right)^2\frac{r_{u_i u'}(t,\x)r_{u_j u'}(t,\x)}{(\sum_{j=1}^{\overline{m}}r_{u_j u'}(t,\x))^2}}{(ce^{-\sum_{j=1}^{\overline{m}}\int_{t}^{t+h}r_{u_j u'}(s,\x)ds}-1)(ce^{-\sum_{j=1}^{\overline{m}}\int_{t}^{t+h}r_{u_j u'}(s,\x)ds}-2)}+o(h).
\end{align}
Plugging \eqref{eqn:cov_negative_multinomial1} into \eqref{eqn:cov_negative_multinomial}, we have
\begin{align}
	\label{eqn:cov_negative_multinomial2}
	&\Cov[\Delta^{\X}_{u_iu'}(t,h),\Delta^{\X}_{u_ju'}(t,h)\mid  \X(t)=\x]\nonumber\\
	&=(x_{u'}^2+x_{u'})\frac{c^2\left(1-e^{-\sum_{j=1}^{\overline{m}}\int_{t}^{t+h}r_{u_j u'}(s,\x)ds}\right)^2\frac{r_{u_i u'}(t,\x)r_{u_j u'}(t,\x)}{(\sum_{j=1}^{\overline{m}}r_{u_j u'}(t,\x))^2}}{(ce^{-\sum_{j=1}^{\overline{m}}\int_{t}^{t+h}r_{u_j u'}(s,\x)ds}-1)(ce^{-\sum_{j=1}^{\overline{m}}\int_{t}^{t+h}r_{u_j u'}(s,\x)ds}-2)}\nonumber\\
	&\quad-x_{u'}^2\frac{c\left(1-e^{-\sum_{j=1}^{\overline{m}}\int_{t}^{t+h}r_{u_j u'}(s,\x)ds}\right)\frac{r_{u_i u'}(t,\x)}{\sum_{j=1}^{\overline{m}}r_{u_j u'}(t,\x)}}{(ce^{-\sum_{j=1}^{\overline{m}}\int_{t}^{t+h}r_{u_j u'}(s,\x)ds}-1)}\nonumber\\
	&\hspace*{1.5cm}\times\frac{c\left(1-e^{-\sum_{j=1}^{\overline{m}}\int_{t}^{t+h}r_{u_j u'}(s,\x)ds}\right)\frac{r_{u_j u'}(t,\x)}{\sum_{j=1}^{\overline{m}}r_{u_j u'}(t,\x)}}{(ce^{-\sum_{j=1}^{\overline{m}}\int_{t}^{t+h}r_{u_j u'}(s,\x)ds}-1)}+o(h).
\end{align}
By Definition \ref{def:infinitesimal_covariance}, for $i,j\in \{1,\ldots,\overline{m}\}$ and $i\neq j$,
\begin{align*}
	\sigma_{u_iu',u_ju'}^{d\X}(t,\x)=0.
\end{align*}
\smallskip

\noindent $(2)$. By \eqref{eqn:negative_multinomial_pdf_modified},
with $\mathbf{y}=(y_0,y_1,\ldots,y_{\overline{m}})$, we have
\begin{align}
	\label{eqn:noise_pdf_negative_multinomial}
	&\mathbb{P}(\{\Delta^{\X}_{u_iu'}(t,h)=k_{i}\}_{i\in \{1,\ldots,\overline{m}\}} \mid \X(t)=\x)\nonumber\\
	&=\int_0^1\frac{\Gamma(x_{u'}+\sum_{i=1}^{\overline{m}}k_{i})}{\Gamma(x_{u'})\prod_{i=1}^{\overline{m}}\Gamma(k_{i}+1)}\left[y_0\right]^{x_{u'}}\prod_{i=1}^{\overline{m}}\left[y_i\right]^{k_{i}}\frac{\Gamma(\sum_{i=0}^{\overline{m}}\alpha_{u_i u'}(t,h,\x) )}{\prod_{i=0}^{\overline{m}}\Gamma(\alpha_{u_i u'}(t,h,\x))}\nonumber\\
	&\quad\quad\quad\times\prod_{i=0}^{\overline{m}}y_i^{\alpha_{u_i u'}(t,h,\x)-1}d\mathbf{y}+o(h)\nonumber\\
	&=\frac{\Gamma(x_{u'}+\sum_{i=1}^{\overline{m}}k_{i})}{\Gamma(x_{u'})\prod_{i=1}^{\overline{m}}\Gamma(k_{i}+1)}\frac{\Gamma(\sum_{i=0}^{\overline{m}}\alpha_{u_i u'}(t,h,\x) )}{\prod_{i=0}^{\overline{m}}\Gamma(\alpha_{u_i u'}(t,h,\x))}\nonumber\\
	&\quad\quad\quad\times\int_0^1\left[y_0\right]^{x_{u'}+\alpha_{u_0 u'}(t,h,\x)-1}\prod_{i=1}^{\overline{m}}\left[y_i\right]^{k_{i}+\alpha_{u_i u'}(t,h,\x)-1}d\mathbf{y}+o(h)\nonumber\\ 
	&=\frac{\Gamma(x_{u'}+\sum_{i=1}^{\overline{m}}k_{i})}{\Gamma(x_{u'})\prod_{i=1}^{\overline{m}}\Gamma(k_{i}+1)}\frac{\Gamma(\sum_{i=0}^{\overline{m}}\alpha_{u_i u'}(t,h,\x) )}{\prod_{i=0}^{\overline{m}}\Gamma(\alpha_{u_i u'}(t,h,\x))}\nonumber\\
	&\quad\quad\quad\times\frac{\Gamma(x_{u'}+\alpha_{u_0 u'}(t,h,\x))\prod_{i=1}^{\overline{m}}\Gamma(k_{i}+\alpha_{u_i u'}(t,h,\x))}{\Gamma(x_{u'}+\sum_{i=1}^{\overline{m}}k_{i}+\sum_{i=0}^{\overline{m}}\alpha_{u_i u'}(t,h,\x))}+o(h)\nonumber\\ &=\frac{\Gamma(x_{u'}+\sum_{i=1}^{\overline{m}}k_{i})}{\Gamma(x_{u'})\prod_{i=1}^{\overline{m}}\Gamma(k_{i}+1)}\frac{\Gamma(c )}{\prod_{i=0}^{\overline{m}}\Gamma(\alpha_{u_i u'}(t,h,\x))}\nonumber\\
	&\quad\quad\quad\times\frac{\Gamma(x_{u'}+\alpha_{u_0 u'}(t,h,\x))\prod_{i=1}^{\overline{m}}\Gamma(k_{i}+\alpha_{u_i u'}(t,h,\x))}{\Gamma(x_{u'}+\sum_{i=1}^{\overline{m}}k_{i}+c)}+o(h).
\end{align}
Recalling that for $i\in \{1,\ldots,\overline{m}\}$
\begin{align*}
	\alpha_{u_i u'}(t,h,\x)
	=&c\left(1-e^{-\sum_{i=1}^{\overline{m}}\int_{t}^{t+h}r_{u_iu'}(s,\x)ds}\right)\frac{r_{u_iu'}(t,\x)}{\sum_{i=1}^{\overline{m}}r_{u_iu'}(t,\x)},
\end{align*}
by Taylor series we have
\begin{align}
	\label{eqn:property0_negative_multinomial}
	\alpha_{u_i u'}(t,h,\x)=cr_{u_iu'}(t,\x) h+o(h).
\end{align}
Hence, for $k_{i}\geq 1$ and $i\in \{1,\ldots,\overline{m}\}$, 
\begin{align}
	\label{eqn:property1_negative_multinomial}
	&\hspace*{-0.1cm}\Gamma(k_i+\alpha_{u_i u'}(t,h,\x))\nonumber\\
	&=(k_i+\alpha_{u_i u'}(t,h,\x)-1)\cdots (2+\alpha_{u_i u'}(t,h,\x)-1)\alpha_{u_i u'}(t,h,\x)\Gamma(\alpha_{u_i u'}(t,h,\x))\nonumber\\
	&=(k_i-1)\cdots (2-1)\cdot\alpha_{u_i u'}(t,h,\x)\cdot\Gamma(\alpha_{u_i u'}(t,h,\x))+o(\alpha_{u_i u'}(t,h,\x))\nonumber\\
	&=\Gamma(k_i)\alpha_{u_i u'}(t,h,\x)\Gamma(\alpha_{u_i u'}(t,h,\x))+o(\alpha_{u_i u'}(t,h,\x)).
\end{align}
Furthermore, we have 
\begin{align}
	\label{eqn:property2_negative_multinomial}
	&\Gamma(x_{u'}+\alpha_{u_0 u'}(t,h,\x))\nonumber\\
	&=\left(x_{u'}-1+c-\sum_{i=1}^{\overline{m}}\alpha_{u_i u'}(t,h,\x)\right)\cdots\left(c-\sum_{i=1}^{\overline{m}}\alpha_{u_i u'}(t,h,\x)\right)\Gamma(\alpha_{u_0 u'}(t,h,\x))\nonumber\\
	&=\left[(x_{u'}-1+c)\cdots c+\mathcal{O}\left(\sum_{i=1}^{\overline{m}}\alpha_{u_i u'}(t,h,\x)\right)\right]\Gamma(\alpha_{u_0 u'}(t,h,\x))\nonumber\\
	&=\left[\frac{\Gamma(x_{u'}+c)}{\Gamma(c)}+\mathcal{O}\left(\sum_{i=1}^{\overline{m}}\alpha_{u_i u'}(t,h,\x)\right)\right]\Gamma(\alpha_{u_0 u'}(t,h,\x)).
\end{align}
Plugging \eqref{eqn:property0_negative_multinomial}, \eqref{eqn:property1_negative_multinomial} and \eqref{eqn:property2_negative_multinomial} into \eqref{eqn:noise_pdf_negative_multinomial}, we can see that
\begin{align*}
	\mathbb{P}(\{\Delta^{\X}_{u_iu'}(t,h)=k_{i}\}_{i\in \{1,\ldots,\overline{m}\}},\; |\overline{\mathcal{S}}|\geq 2  \mid \X(t)=\x)=o(h)
\end{align*}
and
\begin{align*}
	\mathbb{P}(\{\Delta^{\X}_{u_iu'}(t,h)=k_{i}\}_{i\in \{1,\ldots,\overline{m}\}},\; |\overline{\mathcal{S}}|=1  \mid \X(t)=\x)=\sum_{i=1}^{\overline{m}}q_{u_iu'}(t,\x,k_i)h+o(h),
\end{align*}
where $\overline{\mathcal{S}}$ is the set defined in \eqref{eqn:setA_negative_multinomial}, $|\overline{\mathcal{S}}|$ is the cardinality of $\overline{\mathcal{S}}$, and
\begin{align*}
	q_{u_iu'}(t,\x,k_i)
	=c\frac{\Gamma(x_{u'}+\sum_{i=1}^{\overline{m}}k_{i})}{\Gamma(x_{u'})\prod_{i=1}^{\overline{m}}\Gamma(k_{i}+1)}\frac{\Gamma(x_{u'}+c)\Gamma(k_{i})}{\Gamma(x_{u'}+\sum_{i=1}^{\overline{m}}k_{i}+c)}r_{u_iu'}(t,\x).
\end{align*}
\bigskip

\end{document}